\documentclass[12pt]{article} 

\usepackage{amssymb,amsmath,color}
\usepackage{epsfig}
\usepackage{epstopdf}
\usepackage{framed}

\newcommand{\rem}[1]{}
\newtheorem{theorem}{Theorem}

\newtheorem{assumption}[theorem]{Assumption}

\newtheorem{lemma}[theorem]{Lemma}

\newtheorem{remark}[theorem]{Remark}

\numberwithin{theorem}{section}

\def\be{\begin{equation}}
\def\ee{\end{equation}}
\def\bea{\begin{eqnarray}}
\def\eea{\end{eqnarray}}
\def\ba{\begin{array}}
\def\ea{\end{array}}

\def\bx{{\mathbf {x} }}

\let\<\langle
\let \>\rangle

\newcommand{\comment}[1]{\vspace{5 mm}\par \noindent
\marginpar{\textrm{NB!}}
\framebox{\begin{minipage}[c]{ \textwidth}
\tt #1 \end{minipage}}\vspace{5 mm}\par}
\newcommand{\bfi}{{\bf \em}}

\begin{document}
\title{Kinetic models of heterogeneous dissipation}
\author{\vspace{2mm}
\hspace{-0mm}
D. D. Holm$^{1,\,2}$, V. Putkaradze$^{3,4}$, and C. Tronci$^{1,5}\!$\\
\\
{\small\hspace{-10mm} $^1$ \it Department of Mathematics, Imperial College London, London SW7 2AZ, UK}\\
{\small\hspace{-10mm} $^2$ \it Computer and Computational Science Division},  \\ 
 {\small \it Los Alamos National Laboratory,} 
{\small\it Los Alamos, NM, 87545 USA} \\
{\small\hspace{-10mm} $^3$ \it Department of Mathematics, Colorado State University, 
Fort Collins CO 80235 USA} \\
{\small\hspace{-10mm} $^2$ \it Department of Mechanical Engineering,
University of new Mexico, Albuquerque NM 87106 USA} \\
{\small\hspace{-10mm} $^5$\,\it TERA Foundation for Oncological Hadrontherapy,
11 V. Puccini, Novara 28100, Italy}
\\ \\
}

\maketitle

\begin{abstract}\noindent
We suggest kinetic models of dissipation for an ensemble of interacting oriented particles, for example, moving magnetized  particles. This is achieved by  introducing a double bracket dissipation in kinetic equations using an oriented Poisson bracket, and employing the moment method to derive continuum equations for magnetization and density evolution. We show how our continuum equations generalize 
the Debye-H\"uckel equations for attracting round particles, and Landau-Lifshitz-Gilbert equations for spin waves in magnetized media. 
We also show formation of singular solutions that are clumps of aligned particles (\emph{orientons}) starting from random initial conditions. Finally, we  extend our theory to the dissipative motion of self-interacting curves.  
\end{abstract}

\tableofcontents
\section{Introduction}

Many physical systems -- such as plasmas, particle-laden fluids, self assembling molecules \emph{etc.} -- can be understood in terms of collective dynamics of a large number of particles moving in an ambient medium. 
Dissipation plays a crucial role for these systems, in forcing them towards an eventual equilibrium state in the absence of external forces. The dissipation in these \emph{heterogeneous} systems is 
fundamentally different from internal dissipation in homogeneous fluids, which is usually described by diffusion terms in momentum equations (like the Navier-Stokes equations). 
The goal of this paper is to demonstrate a unifying principle for deriving dissipation in systems consisting of particles which are embedded within an ambient fluid, or matrix. The only role of that matrix is to exert friction on individual particles at the microscopic level, and the matrix itself is assumed to have no dynamics of its own. This friction is applied both in configuration space and in momentum space, which leads to the necessity of considering kinetic equations. 

The physical picture we have in mind consists of microscopic magnetized particles moving under the influence of external forces and mutual attraction in a passive ambient continuum. Such systems have attracted much attention recently because of their technological significance. For example, magnetic nanoparticles in this setting have yielded crystalline structures with densities far exceeding that achievable using lithographic procedures  \cite{Sunetal2000,Zengetal2002,BaCoKe2005}.  

Much of the modern modeling of dissipation in systems of moving nano-particles is based on the work of George Gabriel Stokes, who formulated his famous drag law for the resistance of spherical particles moving in a viscous fluid at low Reynolds numbers. Reynolds number is defined as the ratio $R=UL/\nu$ where $U$ is the typical velocity, $L$ is the typical size and $\nu$ is the kinematic viscosity. Reynolds number is physically the ratio between a typical inertial term and viscous force, and $R \ll 1$ means the total dominance of viscous forces. It is commonly assumed that all processes in fluids at micro- and nano-scales are dominated by viscous forces and the Stokes approximation applies. 
The Stokes result states that a  round particle moving through ambient fluid will experience resistance force that is proportional to the velocity of the particle \cite{Ba2000}. Conversely, in the absence of inertia, the velocity of a particle will be proportional to the force applied to it since resistance force and applied force must balance.  This law, that ``force is proportional to velocity'' is sometimes called \emph{Darcy's law}  \cite{Darcy1856}. 

Generalizing this concept to apply to an assembly of a large number of particles, we define the relative density of particles as the number of particles per unit volume, $\rho$. Given the energy of particles in given configuration $E[\rho]$, 
the local velocity of  particles under the force $\mathbf{F}=- \nabla \delta E/\delta \rho$ is $\mathbf{u}=-\mu_{ij} F_j$.   The coefficient of proportionality is called the \emph{mobility matrix} $\mu_{ij}$. For round particles, 
$\mu_{ij}=\mu \delta_{ij}$ where $\delta_{ij}$ is the Kronecker delta (equal to 1 if $i=j$ and $0$ otherwise), so local velocity $\mathbf{u}$ is simply proportional to the local force $\mathbf{F}$, $\mathbf{u}=\mu \mathbf{F}$.
In most cases, the mobility is extended to depend on $\rho$ itself.
In particular here we allow for $\mu$ to be an average value of the dynamical variable, which is given by a convolution $\mu=K*\rho$ with some kernel $K$. The equation of motion for the density $\rho$ is then given by the following conservation law whose general setting can be traced back to Debye and H\"uckel \cite{DeHu1923}
\begin{equation} 
\frac{\partial \rho}{\partial t}=- \mbox{div} \rho \mathbf{u} = 
\mbox{div} \rho \mu \nabla \frac{\delta E}{\delta \rho} \, . 
\label{darcy}
\end{equation} 
 This equation neglects the diffusion of $\rho$ which may occur if the particles also experience Brownian motion.  Note that the rate of change of the energy $E[\rho]$ and an arbitrary functional of density $F[\rho]$ is  given by:  
 \begin{equation} 
 \frac{dE}{dt}=- \int \rho \mu[\rho] \left( \nabla \frac{\delta E}{\delta \rho} \right)^2 \mbox{d} \mathbf{x} \, , 
 \quad 
 \quad 
  \frac{dF}{dt}=- \int \rho \mu[\rho]  \nabla \frac{\delta E}{\delta \rho} \cdot  \nabla \frac{\delta F}{\delta \rho} \mbox{d} \mathbf{x} \, ,
 \label{dissenergy}
 \end{equation} 
which can be seen by direct application of equation (\ref{darcy}). See also \cite{Ot2001,HoPu2005,HoPu2006} for relevant discussions. 

Despite apparent mathematical differences, the work of Gilbert on dissipation in the Landau-Lifshitz equation for spin waves may be recast into the Darcy law form. Indeed, Gilbert \cite{Gi2004}  assumed a dissipation 'force' $F[\mathbf{M}]$ for a set of spins $\mathbf{M}$  to be proportional in the rate of change of $\mathbf{M}$ that we shall denote $\dot{\mathbf{M}}$: 
\begin{equation} 
F_i[\mathbf{M}]= \eta_{ij}  \dot{{M}_j} \, . 
\label{GilbertForce}
\end{equation} 
Gilbert proposed that in the absence of any information on the matrix $\eta_{ij}$ it is prudent to assume simply  $\eta_{ij}=\eta \delta_{ij}$, which ultimately leads to the Landau-Lifshitz-Gilbert dissipation term \cite{Gi2004, LaLi1935} which we write as 
\begin{equation} 
\frac{\partial \mathbf{M}}{\partial t}=\gamma \mathbf{M} \times\mathbf{H}+ 
 \frac{\lambda}{M^2}\mathbf{M} \times\mathbf{M} \times \mathbf{H}
 \, , 
 \label{Gilbert}
\end{equation} 
where $\mathbf{H}$ is the magnetic field, which can be expressed in terms of the external field $\mathbf{H}_{ext}$ and the energy of particle interactions  $E[\mathbf{M}]$ as 
$\mathbf{H}=\mathbf{H}_{ext}+\delta E/\delta \mathbf{M}$ (the variational, or Fr\'echet derivative) and $\gamma, \,\lambda$ are constants. We shall note that alternative phenomenological theories of dissipation exist (Bloch-Brombergen \cite{Bl1950}, COT \cite{CoOlTo1955}) that are not in a generalized Darcy's form.  The Hamiltonian part of this equation  (the first term on the right hand side) has the same dynamics as the rigid body, in terms of angular momentum. The important feature of this model is that it conserves the magnitude of the magnetization vector $|\bf M|$, while dissipating the total energy $\gamma\,\int\bf M\cdot H\,{\rm d}^3\boldsymbol{x}$. 

If we only take into account the dissipative (right hand side) terms in (\ref{Gilbert}), then the rate of change in the energy $E[\mathbf{M}]$ and an arbitrary 
functional of the magnetization $\mathbf{M}$ is given by 
\begin{equation} 
 \frac{dE}{dt}=- \int \left(\mathbf{M} \times  \frac{\delta E}{\delta \mathbf{M}} \right)^2 \mbox{d} \mathbf{x} \, , 
 \quad 
 \quad 
  \frac{dF}{dt}=- \int\left(\mathbf{M} \times  \frac{\delta E}{\delta \mathbf{M}} \right) \cdot \left(\mathbf{M} \times  \frac{\delta F}{\delta \mathbf{M}} \right) \mbox{d} \mathbf{x} \, ,
 \label{dissenergy-Gilbert}
 \end{equation} 
Thus, in essence, both the Stokes drag law and Gilbert dissipation are based on a single mathematical assumption: a `dissipative force' acting on a physical quantity  is proportional to  the rate of change of that physical quantity. The Stokes drag law is valid for mobile particles that interact via collisions; Gilbert's work is developed for particles with the immobile center of mass that can orient themselves and change their magnetic properties. As far as we are aware, the work since Gilbert and Landau-Lifshitz on magnetic dissipation has remained largely phenomenological. A theory of magnetic dissipation for freely moving particles that is applicable to the nano-particle self-assembly still remains to be derived. The challenge is place the phenomenological arguments that have been shown historically to be practical for magnetic dissipation into a systematic theoretical framework. 

This paper derives equations for the dynamics of magnetized particles moving through a passive ambient fluid. Our derivation of the equations motion is \emph{not} based on phenomenological arguments. Equations (\ref{dissenergy}) and 
(\ref{dissenergy-Gilbert}) possess intriguing physical and mathematical similarities, but a direct generalization to moving particles with orientation does not seem feasible. Instead, we proceed by introducing dissipation into the kinetic (Vlasov) equation   describing the motion of particle density $f(p,q,t)$ in the momentum $p$ and coordinate $q$ space. Only that approach can be physically justified. 
  Our starting point is the formulation of the dissipative kinetic theory of gases consisting of arbitrarily shaped microscopic rigid bodies by accurately taking into account the linear and angular momentum of a microscopic particle. The continuum equations of motion for the density and magnetization are then obtained by integrating out the momentum coordinate and truncating the expansion to the first order (a procedure known as `moment truncation' in plasma physics). 
Remarkably, our general framework recovers both the continuum conservation law (\ref{darcy}) for particles without orientation, and the Gilbert dissipation (\ref{Gilbert}) for the oriented particles with the immobile center of mass. 
Moreover in some particular systems, we observe the formations of domains of coherent orientation and we analytically predict the evolution of each domain.

We note that an alternative form of dissipation in continuum media, caused by Brownian motion and expressed as diffusion terms in the Fokker-Planck equations is not suitable for our purposes, as applied to density, orientation or Vlasov form.  The Fokker-Planck equation is a drift-diffusion extension of the Vlasov equation for kinetic dynamics that is widely used as a powerful tool in understanding the physics of certain dissipative phenomena, especially in plasma physics. 
In this regard we mention some important results by Chavanis et al. (see \cite{Chavanis}
and references therein), where a particular form of equation (\ref{darcy}) with $\mu=const.$ has been obtained from a generalized Fokker-Plank treatment
in the simpler case of isotropic interactions. However, informed by recent results in nano-scale self-assembly, we assume that Brownian particle diffusion may be neglected for our purposes. For other experiments with strong diffusive effects, one would seek an alternative theory. 

\section{A dissipative kinetic equation for collisionless systems}
Recently, it was demonstrated  \cite{HoPuTr2007-CR} that the Darcy's conservation law (\ref{darcy}) can be obtained from the introduction of dissipation in the kinetic Vlasov equations, describing the evolution of the density $f(p,q,t)$ in the phase space: 
\begin{equation}
\frac{\partial f}{\partial t}
+
\underbrace{ \left\{f,\frac{\delta H}{\delta f}\right\} }_{\mbox{Inertial part} }
=
\underbrace{ \left\{f,\left\{\mu[f],\frac{\delta E}{\delta f}\right\}\,\right\} }_{\mbox{Dissipation part}}
\, , 
\label{Vlasov-diss-norotation}
\end{equation}
where $H[f]$ is the Vlasov Hamiltonian and ${\delta H}/{\delta f}$ is its variational
(Fr\'echet)
derivative, governing the single particle motion.  The \emph{mobility} (in phase space)  $\mu[f]$ is a modeling choice 
and the energy functional $E[f]$ is usually taken to be the potential part of the Hamiltonian $H$, describing just the self-interactions between the particles in the momentum space. If the inertial part is neglected, then by integrating this equation with respect to the momentum $(p)$ we obtain precisely equation (\ref{darcy}), with $\rho(q,t)=\int f(p,q,t) \mbox{d} p$ and $\mu[\rho]=\int \mu[f(p,q,t)] \mbox{d} p$.  A particular form of  equation (\ref{Vlasov-diss-norotation}) with $\mu[f]=f$ was first suggested by Kandrup \cite{Ka1991} in the context of galactic dynamics, and then analyzed later by Bloch \emph{et al.} \cite{BlBrRa1992, BlKrMaRa1996}. The structure of this equation is consistent with the work of Kaufman
\cite{Ka1984} and Morrison \cite{Mo1984}, where it is shown how dissipation can be introduced
in Hamiltonian systems through a symmetric bracket, instead of the antisymmetric
bracket, which is typical of Hamiltonian dynamics. This work was later extended by Grmela \cite{Gr1984, Gr1993a, Gr1993b}.

We list a series of relevant results that were proved in 
\cite{HoPuTr2007-CR}. 
\begin{lemma}
Equation (\ref{Vlasov-diss-norotation}) preserves  the single-particle
solutions 
\[ 
f=\sum_k w_k(t) \delta\big(p-p_k(t) \big)   \delta\big(q-q_k(t) \big) \, , 
\] 
(with $\dot{w}_k=0$) that are not allowed in other approaches in kinetic dissipation.  
 \end{lemma}

\begin{lemma}
Equation (\ref{Vlasov-diss-norotation}) preserves entropy 
$S=\int f \log f \mbox{d} p \mbox{d}q$ and has infinitely many Casimirs: 
for an arbitrary function $\phi(f)$, the integral $\int \phi(f) \mbox{d} p \mbox{d}q$ is constant, independently of the choice of interaction energy. 
\end{lemma} 
\begin{lemma} 
If the inertial part is discarded, then truncating the moment equations and keeping only the terms dependent on $\rho = \int f(p,q,t) \mbox{d} p$ (zero-th
moment terms)
yields exactly Debye-H\"uckel equation  (\ref{darcy}). 
\end{lemma}

Let us now demonstrate how the dissipation given by (\ref{Vlasov-diss-norotation}) is connected to the Darcy's law.  We neglect the inertial part of (\ref{darcy}) and consider the rate of change for the energy $E[f]$ and  an arbitrary functional $F[f]$ using Kandrup's choice for the mobility $\mu[f]=f$: 
\begin{equation}
\frac{dE}{dt}=- \int \left\{f,\frac{\delta E}{\delta f}\right\} ^2 \mbox{d}\mathbf{x} \, , 
\quad 
\quad 
\frac{dF}{dt}=- \int \left\{f,\frac{\delta E}{\delta f}\right\} \left\{f,\frac{\delta F}{\delta f}\right\} \mbox{d}\mathbf{x} 
\, . 
\label{dissenergy-Vlasov-norotation}
\end{equation} 
This result can also be easily  generalized to mobility functionals having the form $\mu[f]=f\,M[f]$, with $M[f]\geq0$ being a scalar functional of $f$ \cite{HoPuTr2007-CR}. For general mobility functionals $\mu[f]$, the quadratic form given by energy dissipation functional (\ref{dissenergy-Vlasov-norotation}) is not necessarily positive definite. 

Thus, just  as   equation (\ref{darcy}), the dissipation in (\ref{Vlasov-diss-norotation}) enforces the rate of change of an arbitrary functional of the solution to be linear in both $\delta E/\delta f$ and $\delta F/ \delta f$. In particular, energy dissipation is given by a quadratic form of $\delta E/ \delta f$. This formulation in terms of energy dissipation is actually equivalent to saying `force is proportional to velocity', but it has much more physical and mathematical  power, which we are going to exploit in this paper. 

The mathematical equivalence of the dissipation laws (\ref{dissenergy}, \ref{dissenergy-Gilbert}) and (\ref{dissenergy-Vlasov-norotation}) is now clear: 
\begin{assumption} 
The energy dissipation of an arbitrary  functional  $F[f]$ is a quadratic form in the variational derivative of $F[f]$ and the energy $E[f]$ -- see (\ref{FGdist}) below. 
\end{assumption} 
There seems to be no direct path to consistently generalize the evolution equation for moving magnets  to include both density and orientation, somehow combining (\ref{darcy}) and (\ref{Gilbert}). The path we follow is to extend the kinetic equation (\ref{Vlasov-diss-norotation}) to account for orientation interaction and then compute the evolution equations using the moments approach.

\section{Historic digression and mathematical background }
\subsection{Selective decay hypothesis} 
The inspiration for this work originated  in the dissipative bracket structure introduced in Bloch et al. \cite{BlKrMaRa1996},
which in turn was  motivated in part by the double bracket introduced by Vallis, Carnevale, and Young \cite{VaCaYo1989} for incompressible fluid flows.
The dominant idea for  \cite{VaCaYo1989} was the {\bfi selective decay hypothesis}, which arose in turbulence research \cite{MaMo1980} and is consistent with the preservation of coadjoint orbits. 
According to the selective decay hypothesis, energy in strongly nonequilibrium statistical systems tends to decay much faster than certain other ideally conserved properties. In particular, energy decays much faster in such systems than those ``kinematic'' or ``geometric'' properties that would have been preserved in the ideal nondissipative limit {\it independently of the choice of the Hamiltonian}. Examples are Casimir functions for the Lie-Poisson formulations of various ideal fluid models \cite{HoMaRaWe1985}. 

The selective decay hypothesis was inspired by a famous example; namely, that enstrophy decays much more slowly than kinetic energy in 2D incompressible fluid turbulence. Kraichnan \cite{Kr1967} showed that the decay of kinetic energy under the preservation of enstrophy causes dramatic effects in 2D turbulence. Namely, it causes the well known ``inverse cascade'' of kinetic  energy to {\it larger} scales, rather than the ``forward cascade'' of energy to smaller scales that is observed in 3D turbulence! In 2D ideal incompressible fluid flow the enstrophy (the $L^2$ norm of the vorticity) is preserved on coadjoint orbits. That is, enstrophy is a Casimir of the Lie-Poisson bracket in the Hamiltonian formulation of the 2D Euler fluid equations. Vallis et al. \cite{VaCaYo1989} chose a form of dissipation that was expressible as a double Lie-Poisson bracket. This choice of dissipation preserved the enstrophy and thereby enforced the selective decay hypothesis for all 2D incompressible fluid solutions, laminar as well as turbulent. 

Once its dramatic effects were  recognized in 2D turbulence,  selective decay was posited as a governing mechanism in other systems, particularly in statistical behavior of fluid systems with high variability. For example, the slow decay of magnetic helicity was popularly invoked as a possible means of obtaining magnetically confined plasmas \cite{Ta1986}.  Likewise, in geophysical fluid flows, the slow decay of potential vorticity (PV) relative to kinetic energy strongly influences the dynamics of weather and climate patterns much as in the inverse cascade tendency in 2D turbulence. The use of selective decay ideas for PV thinking in meteorology and atmospheric science has become standard practice since the fundamental work in \cite{HoMcRo1985, Yo1987}.

\subsection{Mathematical framework for geometric dissipation}
As explained in \cite{Ot2001}, dissipation of energy $E$ may naturally summon an appropriate metric tensor. In previous work, Holm and Putkaradze \cite{HoPu2007, HoPuTr2007} showed that for any two functionals  $F[ \kappa],G[\kappa]$ of a geometric quantity $\kappa$ a distance between them may be defined via the  Riemannian metric, 
\begin{equation} 
g_\kappa \left(  F \, , \,  E \right) = 
\left\langle  
\Big(\mu[\kappa] \,\diamond\, \frac{\delta F}{\delta f}\Big)
,\, 
\Big( \kappa\,\diamond\,\frac{\delta E}{\delta f}\,\Big)^\sharp
\right\rangle_{\mathfrak{X}^*\times\mathfrak{X}}
  \,. 
\label{FGdist}
\end{equation} 
Here $\langle\,\cdot\,,\,\cdot\, \rangle$ denotes the $L^2$ pairing of vector fields $(\mathfrak{X})$ with their dual one-form densities $(\mathfrak{X}^*)$,  sharp $(\,\cdot\,)^\sharp$ raises the vector index from covariant to contravariant and $\mu[\kappa]$ is the \emph{mobility functional}. The mobility $\mu[\kappa]$ is assumed to satisfy the requirements for (\ref{FGdist}) to be positive definite and symmetric, as discussed in \cite{HoPuTr2007}. The diamond operation $(\diamond)$ in equation (\ref{FGdist}) is the dual of the Lie algebra action, defined as follows. 
Let a vector field $\xi$ act on a vector space $V$ by Lie derivation, so that the Lie algebra action of $\xi$ on any 
element $\kappa\in V$ is given by the Lie derivative,
\[
\xi\!\cdot\kappa=\pounds_{\!\xi}\,\kappa
\,.
\]
The operation dual to the Lie derivative is denoted by $\diamond$ and  defined in terms of the $L^2$ pairings between $\mathfrak{X}$ and $\mathfrak{X}^*$ and between $V$ and $V^*$ as
\begin{equation} 
\Big\langle \zeta\diamond\kappa,\xi 
\Big\rangle_{\mathfrak{X}^*\times\mathfrak{X}}
:=
\Big\langle \zeta,-\pounds_{\!\xi}\,\kappa 
\Big\rangle_{V^*\times V}
\,.
\label{diamonddef} 
\end{equation} 
Given the metric (\ref{FGdist}) and a dissipated energy functional $E[\kappa]$, the time evolution of \emph{arbitrary} functional $F[\kappa]$ is given by \cite{HoPu2007,HoPuTr2007} as
\begin{equation}
\frac{dF}{dt}=\{\!\{\,F\,,\,E\,\}\!\} [\kappa]
:=-\,g_\kappa \left( F, E \right)= 
-
\left\langle  
\Big(\mu[\kappa] \,\diamond\, \frac{\delta E}{\delta \kappa}\Big)
,\, 
\Big(\kappa\,\diamond\,\frac{\delta F}{\delta \kappa}\,\Big)^\sharp
\right\rangle_{\mathfrak{X}^*\times\mathfrak{X}}
\,,
\label{bracket}
\end{equation}
which specifies the dynamics of any functional $F[\kappa]$, given the  the energy dependence  $E[\kappa]$. The bracket $\{\!\{\,F\,,\,E\,\}\!\}$ is shown to satisfy the Leibnitz product-rule property for a suitable class of mobility functionals $\mu[\kappa]$ in \cite{HoPu2007,HoPuTr2007}. Eq. (\ref{bracket}) and positivity of $g_\kappa( E,E)$ imply that the energy $E$ decays in time until it eventually reaches a critical point, $\delta E/\delta \kappa=0$. 

\begin{remark}
For densities (dual to functions in the $L^2$ pairing), the Lie derivative is the divergence and its dual operation is (minus) the gradient. Thus, for densities the symbol diamond $(\,\diamond\,)$ is replaced by gradient $(\,\nabla\,)$ in the metric defined in Eq. (\ref{bracket}).
\end{remark}

\section{Kinetic equation for anisotropic particles and  Landau-Lifshitz equations}
 We now turn
to the problem of a system composed of immobile particles of arbitrary 
shape. The orientation $\mathbf{m}$ of particles now appears as an 
additional degree of freedom, but the particle momentum $\mathbf{p}$ is going to be neglected in this section since centers of mass for the particles are assumed to be immobile.  In our approach
we consider orientation as given by the angular momentum of each particle.
Thus, instead of considering unit vectors, we consider angular momentum vectors
and each particle is treated as a rigid body. Let us consider the non-dissipative
case first. The evolution equation for the  distribution function $\varphi({\boldsymbol{x},\boldsymbol{m}},t)$ in this  approach is written in the conservation form in the  $\mathbf {U}=(\mathbf{U_x},\mathbf{U_m})$  on the  $\left(\boldsymbol{x},\,\boldsymbol{m}\right)$ space
\begin{equation*}
\frac{\partial \varphi}{\partial t}=
-\text{\rm\large div}_{\left(\boldsymbol{x},\,\boldsymbol{m}\right)}
\left(\varphi\,{\bf U}\right)
=-\text{\rm\large$\nabla$}_{\boldsymbol{x}}\cdot
\left(\varphi\,{\bf U}_{\boldsymbol{x}}\right)-
\frac{\partial}{\partial\boldsymbol{m}}\cdot
\left(\varphi\,{\bf U}_{\boldsymbol{m}}\right)
\, . 
\end{equation*}
As we mentioned, in this section we set $\mathbf{U_x=0}$ since the centers of mass for the particles are assumed to be stationary. 
Since the angular momentum for a rigid body evolves according to the Hamiltonian motion
$
\dot{\boldsymbol{m}}=
\boldsymbol{m}\times\nabla_{\!\boldsymbol{m}} {h}
={\bf U}_{\boldsymbol{m}}
$, then we can write our kinetic equation as
\begin{equation*}
\frac{\partial \varphi}{\partial t}+\left[\varphi,\,\frac{\delta H}{\delta \varphi}\right]=0
\,.
\end{equation*}
where we have used the definition of
the rigid body bracket $\left[\varphi,\,h\right]:=\boldsymbol{m}\cdot\nabla_{\!\boldsymbol{m}}\varphi\times\nabla_{\!\boldsymbol{m}}h$
and the global Hamiltonian is written as $H[\varphi]=\int\!\!\!\int\!\varphi(\boldsymbol{x},\boldsymbol{m})\,h(\boldsymbol{x},\boldsymbol{m})\,{\rm d}^3\boldsymbol{x}\,{\rm d}^3\boldsymbol{m}$. Now that we have obtained the
conservative dynamics in the bracket form, we are motivated to insert the dissipation as the double bracket form of (\ref{Vlasov-diss-norotation}) as follows: 
\begin{equation}
\frac{\partial \varphi}{\partial t}
+
\underbrace{\left[\varphi,\frac{\delta H}{\delta \varphi}\right]}_{\mbox{Inertia}}
=
\underbrace{\left[\varphi,\left[\mu[\varphi],\frac{\delta E}{\delta \varphi}\right]\,\right]}_{\mbox{Dissipation}} 
\, . 
\label{Kinetic-rotation-notmoving}
\end{equation}
In this context, it is reasonable to take
the moments with respect to {\bfi m} and in particular we want the equation for the magnetization
density ${\bf M}(\boldsymbol{x},t):=\int\!\boldsymbol{m}\,\varphi\,{\rm d}^3\boldsymbol{m}$, when
$H=E$. We now introduce the assumption that the single particle energy $h$
is linear in {\bf m} (it is easy to see that this assumption recovers, for
example, the expression of the potential energy of a magnetic moment in a magnetic field).  

Again, let us consider only the dissipation terms 
in (\ref{Kinetic-rotation-notmoving}), \emph{i.e.}, the term on the right hand side  and use Kandrup's mobility 
$\mu[\varphi]=\varphi$. Then, we get the dissipation of energy $E[\varphi]$ as well as an evolution of an arbitrary functional $F[\varphi]$ in the quadratic  form analogous to (\ref{dissenergy-Vlasov-norotation}): 
\begin{equation}
\frac{dE}{dt}=- \int \left[ \varphi,\frac{\delta E}{\delta \varphi}\right] ^2 \mbox{d}\mathbf{x} \, , 
\quad 
\quad 
\frac{dF}{dt}=- \int \left[ \varphi,\frac{\delta E}{\delta \varphi}\right] \left[ \varphi,\frac{\delta F}{\delta \varphi}\right] \mbox{d}\mathbf{x} 
\, . 
\label{dissenergy-Vlasov-notmoving}
\end{equation} 
The extension to more general mobility functional is also possible, but the quadratic form describing the dissipation will not be  symmetric except for some particular choices for the mobility. 

Now, using the method of moments we compute  the equation for the magnetization density $\mathbf{M}$ as
\begin{equation*}
\frac{\partial \bf M}{\partial t}=
{\bf M}\times\frac{\delta H}{\delta \bf M}+{\bf M}\times\boldsymbol\mu[{\bf
M}]\times\frac{\delta E}{\delta \bf M}
\,.
\end{equation*}
recovering exactly  the Landau-Lifshitz-Gilbert equation (LLG) for
magnetization dynamics (\ref{Gilbert}) for $\boldsymbol{\mu}[\mathbf{M}]=\mathbf{M}$, which occurs for  Kandrup's original choice of mobility, $\mu[f]=f$. 
\rem{ 
Interestingly, the linearity assumption in the derivation of this equation
is in agreement with the expression for potential energy for the alignment of the magnetic moment $\boldsymbol{m}$ with the field creating $\bf B$, which
is given by $\Phi=\boldsymbol{m}\cdot\bf B$.
} 

It is interesting to note  that similar ideas for the extensions of Gilbert dissipation were recently suggested phenomenologically in \cite{TuSl2007}.

\section{Dissipative kinetic equation for moving particles with orientation} 
The kinetic approach for immobile oriented particles can be generalized to moving oriented particles. We are using the energy dissipation approach by generalizing the formula (\ref{dissenergy-Vlasov-norotation}) to more general Poisson bracket that involves both density and orientation. 
Interestingly enough, this Poisson bracket has been derived 25 years ago  in the \cite{GiHoKu1982, GiHoKu1983}, but we believe that we are the first to apply this bracket to dissipation in this setting. This modified Poisson bracket $\{ \cdot \, , \, \cdot \}_1$ is defined as
\begin{equation} 
\left\{ f, \frac{\delta E}{\delta f} \right\}_1= \left\{ f, \frac{\delta E}{\delta f} \right\} +  \left[ f, \frac{\delta E}{\delta f} \right] \, , 
\end{equation} 
that is, it is a sum of a canonical Poisson bracket $\{ \cdot \, , \, \cdot \}$ and the Rigid body bracket $[ \cdot \, , \, \cdot ]$. For $g$ belonging to the dual of  a general Lie Algebra $\mathfrak{g}$  (not necessarily $so(3)^*$), we can compute elements $\delta E/\delta g \in \mathfrak{g}$ and 
$\delta F/\delta g \in \mathfrak{g}$. We thus define a generalized rigid body bracket as 
\begin{equation} 
\left[ 
E \, , \, F
\right] 
= 
\left< 
g 
\, , \,
\left[ 
\frac{\delta E}{\delta g}
\, , \, 
\frac{\delta F}{\delta g}
\right]_{\!LA\,}
\right> 
\, . 
\end{equation} 
Here, we have used $[ \, , \, ]_{LA}$ -- the Lie algebra bracket in $\mathfrak{g}$. 
For this choice of the new bracket  the quadratic form describing the dissipation (\ref{dissenergy-Vlasov-norotation})  generalizes naturally to the anisotropic
case as follows: 
\begin{equation} 
\frac{dF}{dt}=- \int \left\{\mu[f],\frac{\delta E}{\delta f}\right\}_{\!1} \left\{f,\frac{\delta F}{\delta f}\right\}_{\!1} \mbox{d}\mathbf{x} 
\, . 
\label{dissenergy-Vlasov-orientation}
 \end{equation} 
Taking the moments of the anisotropic kinetic Vlasov equation following from (\ref{dissenergy-Vlasov-orientation})  leads to a new kind of continuum dynamics naturally generalizing equation (\ref{darcy}) to the case of anisotropic interaction. To simplify the formulas,  we neglect the inertia of the particles. As mentioned in the introduction, this is a common assumption for friction dominated systems. This assumption means neglecting the term $\{f,\,\delta H/\delta f\}$ in the left hand side of (\ref{Vlasov-diss-norotation}). Thus we are left with the following equation which will form the core of further analysis,
\begin{equation}
\frac{\partial f}{\partial t}
=
\left\{f,\left\{\mu[f],\frac{\delta E}{\delta f}\right\}_{\!1}\,\right\}_{\!1}
\,.
\label{Vlasov-diss-orientation}
\end{equation}
The remainder of this section follows the reasonong in \cite{HoPuTr2007}, where the reader may find many more technical details and exposition of some related problems.

\subsection{Landau-Lifshitz through the Smoluchowski approach}
When dealing with kinetic theories, the kinetic moments are a well established
tool for deriving simplified continuum descriptions.
However, in the case of anisotropic interactions, the kinetic moments $A_n=\int
p^n\,f(q,p,g)\,{\rm d}p$
depend  not only on the position coordinate, but also on the 
single-particle Lie algebra coordinate $g$. Thus, the moment equations differ from those found in the isotropic case, since they must include the terms which take into account the anisotropic behavior.

Moment dynamics has a Hamiltonian structure which is justified by the
theorem that taking moments is a Poisson map. In fact, their dynamics is governed by
the Kupershmidt-Manin bracket \cite{GiHoTr2005,GiHoTr07}. This process can
be transferred to the double bracket formulation as follows.
As in the Kupershmidt-Manin approach, the moments are dual to the variables $\beta_n(q,g)$, which are introduced by expanding the Hamiltonian function
$h(q,p,g)$ as $h(q,p,g)=p^n\,\beta_n(q,g)$. The {\bfi Lie algebra action} is given by
\[
\beta_n\cdot  f=\left\{f,p^n\beta_n\right\}_1
\]
The {\bfi dual action} is defined by a  star ($\star$)  operator which is the symplectic equivalent of the diamond ($\diamond$) operation, as follows,
\begin{align*}
\Big\langle f\,\text{\Large$\star$}_{n}\, k,\,\beta_n\Big\rangle:=
\Big\langle f,\, \beta_n\, k\Big\rangle &=
\Big\langle f\!\star k\,,\, p^n \beta_n(q,g) \Big\rangle =
 \left\langle \int p^n\,\{f, k\}_{1}\,dp\,,\,\beta_n \right\rangle
 \, .
\end{align*}
Thus,  the {\bfi star operator} is defined  explicitly for $k=p^m\alpha_m$ as 
\begin{align*}
f\,\,\text{\Large$\star$}_{n}\,k
\,=&\,
{\sf ad}_{\alpha_m}^*A_{m+n-1}
+
\left\langle g,\left[\frac{\partial A_{m+n}}{\partial g},\frac{\partial \alpha_m}{\partial g}\right]\right\rangle \, . 
\end{align*}
where the ${\sf ad}^*$ operator is the Kupershmidt-Manin Hamiltonian operator, see \cite{GiHoTr2005,GiHoTr07}.
We  introduce the {\bfi dissipative bracket} by 
\begin{equation} 
\dot{F}=\{\{F,E\}\}=-\left\langle \mu[f]\,\text{\large$\star$}_{n}\,\frac{\delta E}{\partial f}\,,\,f\,\text{\large$\star$}_{n}\,\frac{\delta F}{\partial f} \right\rangle \, . 
\label{dissbracketSmol}
\end{equation} 
At the simplest level we truncate this sum, by taking $n=0$. For higher order approximations, giving highly complex formulas, see \cite{HoPuTr2007}. By using this evolution equation for an arbitrary functional $F$, the rate of change for zero-th moment  $A_0$ is found to be
\[
\frac{\partial A_0}{\partial t}=
\left\langle g,\left[\frac{\partial A_0}{\partial g},\frac{\partial \gamma_0}{\partial g}\right]\right\rangle  
\]
where one defines $\gamma_{0}:=\mu[f]\star_{\,0}\delta E/\delta f$, which is explicitly given by
\begin{align*}
\gamma_0&= 
\left\langle g,\left[\frac{\partial \mu_0}{\partial g},\frac{\partial \beta_0}{\partial g}\right]\right\rangle
\,.
\end{align*} 
Upon introducing the notation $\{\cdot,\,\cdot\}$ for the Lie-Poisson
bracket on the Lie algebra $\mathfrak{g}$
\[
\left\{k,\,h\right\}:=
\left\langle g,\left[\frac{\partial k}{\partial g},\frac{\partial h}{\partial
g}\right]\right\rangle
\,,
\]
we may write the equation for $\varphi=A_0$ in the double bracket form,
\begin{equation}
\frac{\partial \varphi}{\partial t}
=
\left\{\varphi,\left\{\mu[\varphi],\frac{\delta E}{\delta \varphi}\right\}\,\right\}
\, . 
\end{equation}
This is the purely dissipative moment  equation corresponding to the Vlasov equation for the oriented particles (\ref{Vlasov-diss-orientation}). It is straightforward to see that
the inclusion of the inertial part such as in equation (\ref{Vlasov-diss-norotation})
yields a moment equation of exactly the same form as equation (\ref{Kinetic-rotation-notmoving})
\begin{equation}
\frac{\partial \varphi}{\partial t}
+
\left\{\varphi,\frac{\delta H}{\delta \varphi}\right\}
=
\left\{\varphi,\left\{\mu[\varphi],\frac{\delta E}{\delta \varphi}\right\}\,\right\}
\,. 
\label{Kinetic-rotation-notmoving2}
\end{equation}
We have now shown how to derive equation (\ref{Kinetic-rotation-notmoving})
from the kinetic approach by taking the zero-th moment. Next, one 
could  continue by integrating over the microscopic magnetization  $\boldsymbol{m}$ and thereby obtain the 
Landau-Lisfshitz-Gilbert equation (\ref{Gilbert}). 

In the particular case $g=\boldsymbol{m}\in so(3)$ that is of interest to us, the Lie-Poisson bracket reduces to the rigid body bracket $\left\{\varphi,\,h\right\}:=\boldsymbol{m}
\cdot\nabla_{\!\boldsymbol{m}}\varphi\times\nabla_{\!\boldsymbol{m}}h$. The
global Hamiltonian is written as $H[\varphi]=\int\!\!\!
\int\!\varphi(q,\boldsymbol{m})\,h(q,\boldsymbol{m})\,
{\rm d}q\,{\rm d}^3\boldsymbol{m}$ and an analogous expression
holds for the dissipation energy $E[\varphi]$. The generalization to higher
spatial dimensions is straightforward, since there are no derivatives in spatial coordinates in the equation (\ref{Kinetic-rotation-notmoving2}).

We next take moments with respect to $\boldsymbol{m}$. The particular physical case of  magnetization 
density ${\bf M}(\boldsymbol{x},t):=\int\!\boldsymbol{m}\,\varphi\,{\rm d}^3\boldsymbol{m}$ requires  $H=E$. We assume that the single particle energy $h$ is linear in $\mathbf{m}$ since this assumption recovers the expression of the potential energy of a magnetic moment in a magnetic field.  We consider only the dissipation terms 
in (\ref{Kinetic-rotation-notmoving}), \emph{i.e.}, we drop the second term term on the left  hand side of that equation. Then,  we obtain the  rate of change of the energy $E[\varphi]$, as well as of an arbitrary functional $F[\varphi]$ in the quadratic  form analogous to (\ref{dissenergy-Vlasov-norotation}): 
\begin{equation}
\frac{dE}{dt}=- \int \left[ \mu[f],\frac{\delta E}{\delta f}\right] ^2 \mbox{d}\mathbf{x} \, , 
\quad 
\quad 
\frac{dF}{dt}=- \int \left[ \mu[f],\frac{\delta E}{\delta f}\right] \left[ f,\frac{\delta F}{\delta f}\right] \mbox{d}\mathbf{x} 
\, . 
\end{equation} 
The method of moments gives  the equation for the magnetization density $\mathbf{M}$
\begin{equation*}
\frac{\partial \bf M}{\partial t}=
{\bf M}\times\frac{\delta H}{\delta \bf M}+{\bf M}\times\boldsymbol\mu[{\bf
M}]\times\frac{\delta E}{\delta \bf M}
\, , 
\end{equation*}
which recovers exactly  the Landau-Lifshitz-Gilbert equation (LLG) for
magnetization dynamics (\ref{Gilbert}) for $\boldsymbol{\mu}[\mathbf{M}]=\mathbf{M}$, which occurs for Kandrup's original choice of mobility $\mu[f]=f$. 

\rem{ 
Interestingly, the linearity assumption in the derivation of this equation
is in agreement with the expression of potential energy for the alignment of the magnetic moment $\boldsymbol{m}$ with the field creating $\bf B$, which
is given by $\Phi=\boldsymbol{m}\cdot\bf B$.
} 

\subsection{Dissipative moment dynamics: an alternative treatment}

\rem{
To derive the moment dynamics with orientation dependence, 
we  follow the same steps as in the previous section, beginning by introducing  the quantities
\[
A_n(q,g):=\!\int p^n\,f(q,p,g)\,dp
\,\,\,\quad\text{with}\quad
g\in\mathfrak{g}^*
\,.
\]
One may find the entire hierarchy of equations for these moment quantities and then integrate over $g$ in order to find the equations for the mass density $\rho(q):=\int \!A_0(q,g)\,dg$ and the continuum charge density $G(q)=\int\! g\,A_0(q,g)\,dg$. Without the integration over $g$, such an approach would yield the Smoluchowski approximation for the density $A_0(q,g)$, usually denoted by $\rho(q,g)$. 

This approach is followed in the Sec.~\ref{sec:Smoluchowski}, where the dynamics of $\rho(q,g)$
is presented explicitly. 
}

In this section, we  extend the Kupershidt-Manin approach as in \cite{GiHoKu1982, GiHoKu1983} to generate the dynamics of moments with respect to both momentum $p$ and charge $g$. The main technical complication is that the Lie algebras of physical interest (such as $\mathfrak{so}(3)$) are not one-dimensional and in general are not Abelian. Thus, in the general case one needs to use a multi-index notation as in \cite{Ku1987,GiHoTr2005}. 
We introduce multi-indices $\sigma:=(\sigma_1, \sigma_2,\,\dots,\sigma_N)$,
with $\sigma_i\geq0$, and define $g^\sigma:=g_1^{\sigma_1}\!\!\dots g_N^{\sigma_N}$, where $N={\rm dim}(\mathfrak{g})$. Then, the moments are expressed as
\[
A_{n\,\sigma}(q)\,:=\int p^n\,g^\sigma\,f(q,p,g)\,dp\,dg
\,.
\]
This multi-dimensional treatment encumbers the calculations with technical compllications. For the purposes of this section, we are primarily interested in the equations for $\rho$ and $G$, so we restrict our considerations only to the moments of the form
\[
A_{n,\nu}=\int p^n\,g_\nu\,f(q,p,g)\,dp\,dg
\qquad \nu=0,1,\dots,N \, . 
\]
Here we define $g\,_0=1$ and $\,g_a=\langle\, g,\,\mathbf{e}_a\rangle$ where $\mathbf{e}_a$ is a basis of the Lie algebra and $\langle\, g_b\mathbf{e}^b,\,\mathbf{e}_a\rangle=g_a\in\mathbb{R}$ represents the result of the pairing $\langle\, \cdot\,,\,\cdot\,\rangle$ between an element of the Lie algebra basis and an element of the dual Lie algebra.
We write the single particle Hamiltonian as $h=\delta H/\delta
f=p^n g_\nu\,\delta H/\delta A_{n,\nu}=:p^n g_\nu \,\beta_n^\nu(q)$, which means  that we employ the following assumption. 

\medskip

\begin{assumption} \label{linassump}
The single-particle Hamiltonian $h=\delta H/\delta f$ is linear in $g$ and can be expressed as 
\[ h(q,p,g)= p^n\,\psi_n(q)+p^n\!\left\langle g,\,\overline\psi_n(q) \right\rangle \, ,
\] 
where $\psi_n(q)\in \mathbb{R}$ is a real scalar function and $\overline\psi_n(q)\in \mathbb{R}\times\mathfrak{g}$ is a real Lie-algebra-valued function. 
\end{assumption}


\medskip

\paragraph{Dual Lie algebra action.}
The action of $\beta_n^\nu$ on $f$ is defined as
\[
\beta_n^{\,\nu}\cdot f=\big\{p^n g_\nu\,\beta_n^{\,\nu},\,f\big\}_{\!1}
\qquad\text{ (no sum)} \, . 
\]
The dual of this action is denoted by $(\star_{n, \nu})$. It may be  computed analogously to (\ref{diamonddef}) and  found to be  
\begin{align*}
f\,\,\text{\Large$\star$}_{n,\nu}\,k
\,=&\,
\iint  p^n g_\nu\,\big\{f,\,k\big\}_{\!1}\,{\rm d}p \,{\rm d}g\\
\,=&\,
\int g_\nu \, g_\sigma \,{\sf ad}_{\alpha_m^{\sigma}}^*A_{m+n-1} \,{\rm d}g
+
\int g_\nu
\left\langle g,\left[\frac{\partial A_{m+n}}{\partial g},\frac{\partial (g_\sigma\,\alpha_m^\sigma)}{\partial g}\right]\right\rangle{\rm d}g
\\
\,=&\,\,
{\sf ad}_{\alpha_m^{\sigma}}^* \int\! g_\nu \, g_\sigma \,A_{m+n-1} \,{\rm d}g
+
\int g_\nu
\left\langle g,\left[\frac{\partial A_{m+n}}{\partial g},\frac{\partial (g_a\,\alpha_m^a)}{\partial g}\right]\right\rangle{\rm d}g 
\, . 
\end{align*}
Here, $k=p^m\,g_\sigma\,\alpha_m^\sigma(q)$ and we have used the definition of the moment 
\[
A_n(q,g)=\int p^n\,f(q,p,g)\,dp
\,.\]

\paragraph{Evolution equation.}
Having characterized the dual Lie algebra action, we
may now write the evolution equation for an arbitrary functional $F$ in terms of the dissipative bracket as follows: 
\begin{equation} 
\dot{F}= \{\!\{\,F\,,\,E\,\}\!\}=-\left\langle\!\!\!\left\langle \left(\mu[f]\,\,\text{\Large$\star$}_{n,\nu}\,\,\frac{\delta E}{\partial f}\right)^\sharp\!,\,f\,\,\text{\Large$\star$}_{n,\nu}\,\,\frac{\delta F}{\partial f} \right\rangle\!\!\!\right\rangle
\,,
\label{dissbracket1} 
\end{equation}
where the pairing $\left\langle\!\!\!\left\langle\,\cdot\,,\cdot\,\right\rangle\!\!\!\right\rangle$
is given by integration over the spatial coordinate $q$.
Now we fix $m=0$, $n=1$.  The equation for the evolution of $F=A_{0,\lambda}:=\int g_\lambda
\,A_0 \,{\rm d}g{\rm d}p$ is found from (\ref{dissbracket1}) to be
\begin{align}
\frac{\partial A_{\,0,\lambda}}{\partial t}
\nonumber
=&\,
\frac{\partial}{\partial q}\left({\gamma_{1,\nu}^\sharp}\int \!g_\nu \, g_\lambda \,A_0 \,{\rm d}g\right)
\\
&+
\int g_\lambda \left\langle g,\left(\,
\left[\frac{\partial A_{1}}{\partial g},\frac{\partial (g_a\,{\gamma}_{1,a}^{\,\,\sharp})}{\partial g}\right]+\left[\frac{\partial A_{0}}{\partial g},\frac{\partial (g_a\,{\gamma}_{0,a}^{\,\,\sharp})}{\partial g}\right]\,\right)\right\rangle{\rm d}g \, , 
\label{moments1}
\end{align}
where we have defined the analogues of Darcy's velocities: 
\begin{align*}
\gamma_{0,\nu}:=\,\mu[f]\,\,\text{\Large$\star$}_{0,\nu}\,\,\frac{\delta E}{\delta f}
&=
\int \!g_\nu
\left\langle g,\left[\frac{\partial \widetilde\mu_{k}}{\partial g},\frac{\partial (g_a\,\beta_k^a)}{\partial g}\right]\right\rangle{\rm d}g
=\int \!g_\nu
\left\langle g,\left[\frac{\partial \widetilde\mu_{0}}{\partial g},\frac{\partial (g_a\,\beta_0^a)}{\partial g}\right]\right\rangle{\rm d}g
\end{align*}
and
\begin{align*}
\gamma_{1,\nu}:=\,\mu[f]\,\,\text{\Large$\star$}_{1,\nu}\,\,\frac{\delta E}{\delta f}
&=\,
\frac{\,\,\partial \beta_0^{\,\sigma}}{\partial q}
\int \!g_\nu \, g_\sigma \,\widetilde\mu_{0} \,{\rm d}g
+
\int g_\nu
\left\langle g,\left[\frac{\partial \widetilde\mu_{1}}{\partial g},\frac{\partial (g_a\,\beta_0^a)}{\partial g}\right]\right\rangle{\rm d}g
\,.
\end{align*}
with $\widetilde{\mu}_k:=\int p^k\,f\,{\rm d}p$. 
Here we have assumed that the energy functional $E$ depends only on $A_{0,\lambda}$
(recall that $\beta_n^\lambda:=\delta E/\delta A_{n,\lambda}$), so that we
may fix $k=0$ in the second line. At this point, we further simplify the treatment
by discarding all terms in $\gamma_{1,a}$, that is we truncate the summations
in equation (\ref{moments1}) to consider only terms in $\gamma_{0,0}$, $\gamma_{0,a}$ and $\gamma_{1,0}$. With this simplification,  equation (\ref{moments1})
becomes
\begin{align}
\frac{\partial A_{\,0,\lambda}}{\partial t}
=&\,
\frac{\partial}{\partial q}\left({\gamma_{1,0}}\int  g_\lambda \,A_0 \,{\rm d}g\right)
+
\int g_\lambda \left\langle g,\left(\,
\left[\frac{\partial A_{0}}{\partial g},\frac{\partial (g_a\,{\gamma}_{0,a}^{\,\,\sharp})}{\partial g}\right]\,\right)\right\rangle{\rm d}g \, , 
\label{moments-simplified}
\end{align}
and the expression for $\gamma_{1,0}$ is
\begin{align*}
\gamma_{1,0}:\!&=\,\mu[f]\,\,\text{\Large$\star$}_{1,0}\,\,\frac{\delta E}{\delta f}
=\,
\frac{\,\,\partial \beta_0^{\,\sigma}}{\partial q}
\int  g_\sigma \,\widetilde\mu_{0} \,{\rm d}g
\,.
\end{align*}
Let us now define the \emph{dynamic quantities} 
\begin{eqnarray*}
\rho&=&\int f \,{\rm d}g\,{\rm d}p
\,,
\qquad
\quad\,\,
G\,\,\,\,=\,\,\,\int g\, f\,{\rm d}g\,{\rm d}p
\,.
\end{eqnarray*}
and corresponding  \emph{mobilities} 
\begin{eqnarray*}
\mu_\rho&=&\int \mu[f] \,{\rm d}g\,{\rm d}p
\,,
\qquad
\quad
\mu_G\,\,\,\,=\,\,\,\int g\, \mu[f]\,{\rm d}g\,{\rm d}p
\,.
\end{eqnarray*}
In terms of these quantities, we may write the following 

\begin{theorem}\label{momeqns-thm-simplified}
The moment equations for $\rho$ and $G$ are given by
\begin{align}
\frac{\partial \rho}{\partial t}=&\,\,
\frac{\partial}{\partial q}\Bigg(\rho\,\,
\bigg(
\mu_\rho\,\, \frac{\partial}{\partial q}\frac{\delta E}{\delta \rho}
+
\bigg\langle \mu_G, \,\frac{\partial}{\partial q}\frac{\delta E}{\delta G}\bigg\rangle
\bigg)
\,\,
\Bigg)
\label{rhogen-simplified}
\end{align}
and
\begin{align}
\frac{\partial G}{\partial t}=&\,\,
\frac{\partial}{\partial q}\Bigg(G\left(\mu_\rho\,\, \frac{\partial}{\partial q}\frac{\delta E}{\delta \rho}+
\left\langle \mu_G, \,\frac{\partial}{\partial q}\frac{\delta E}{\delta G}\right\rangle\right) \Bigg)
+ {\rm ad}^*_{\left(\!{\rm ad}^*_{\frac{\delta E}{\delta G}} \mu_G\!\right)^{\!\!\sharp}}\,G
\label{Ggen-simplified}
\,.
\end{align}
\end{theorem}

\begin{remark}
Similar equations (under the name of Geometric Order Parameter equations) were derived via a general geometric considerations  in \cite{HoPu2007}.
The equations presented here reduce to those for the case of commutative
Lie algebras.
\end{remark}

\begin{remark}
Although  in this section the underlying space where particles undergo their motion is assumed to be one dimensional, the extension
to higher spatial dimensions is trivial: since the 
Lie derivative of a density is given by the divergence term in spatial coordinate ${\bf q}$, we need to substitute div and grad instead of one-dimensional derivatives with respect to ${\bf q}$. See (\ref{rodrho},\ref{DarcyGilbert})  immediately below for an explicitly computed particular case. 
\end{remark}

\section{Self-organization dynamics of moving magnets}
\begin{figure}[htbp]
\begin{center}
\includegraphics[height=6cm]{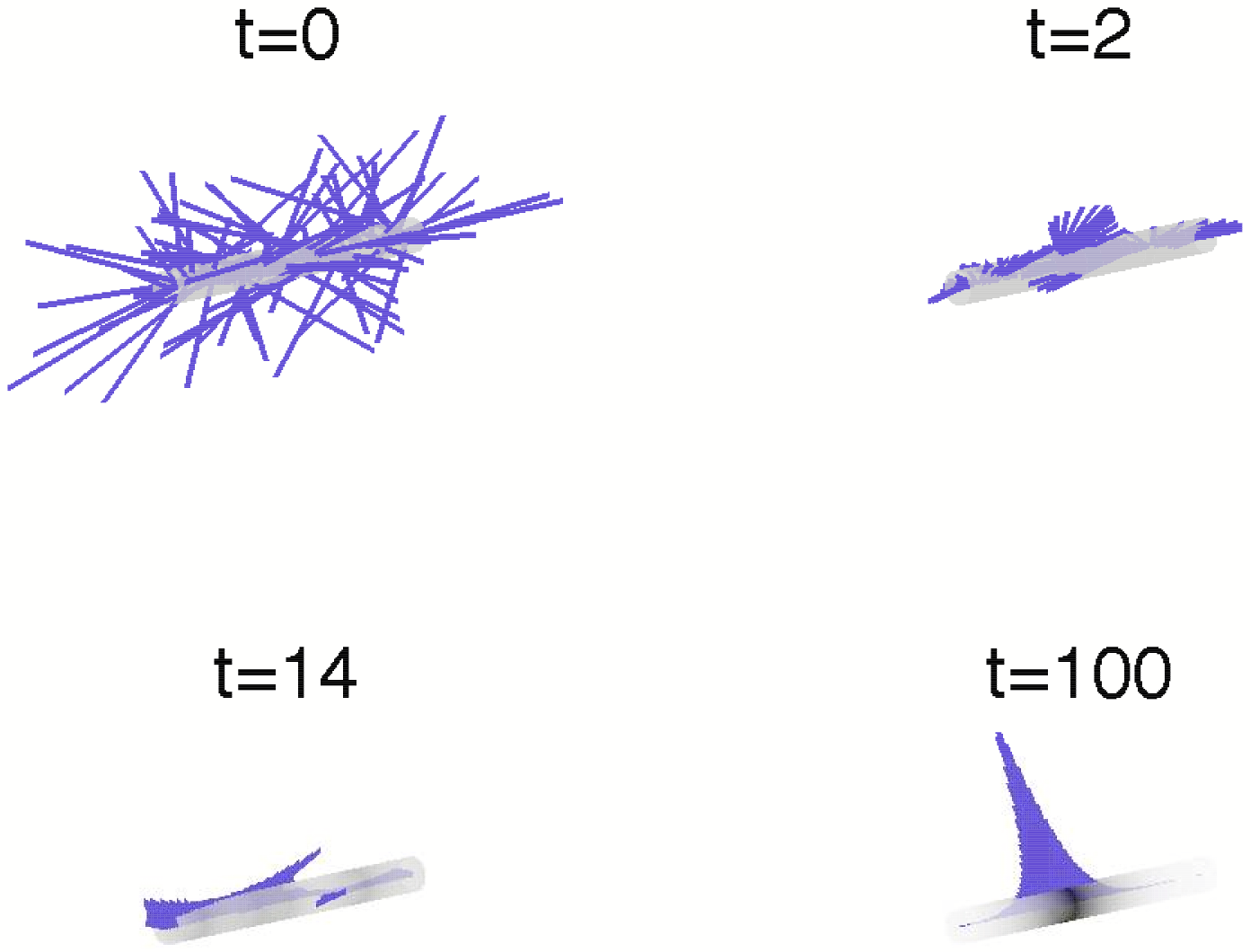}
\includegraphics[height=6cm]{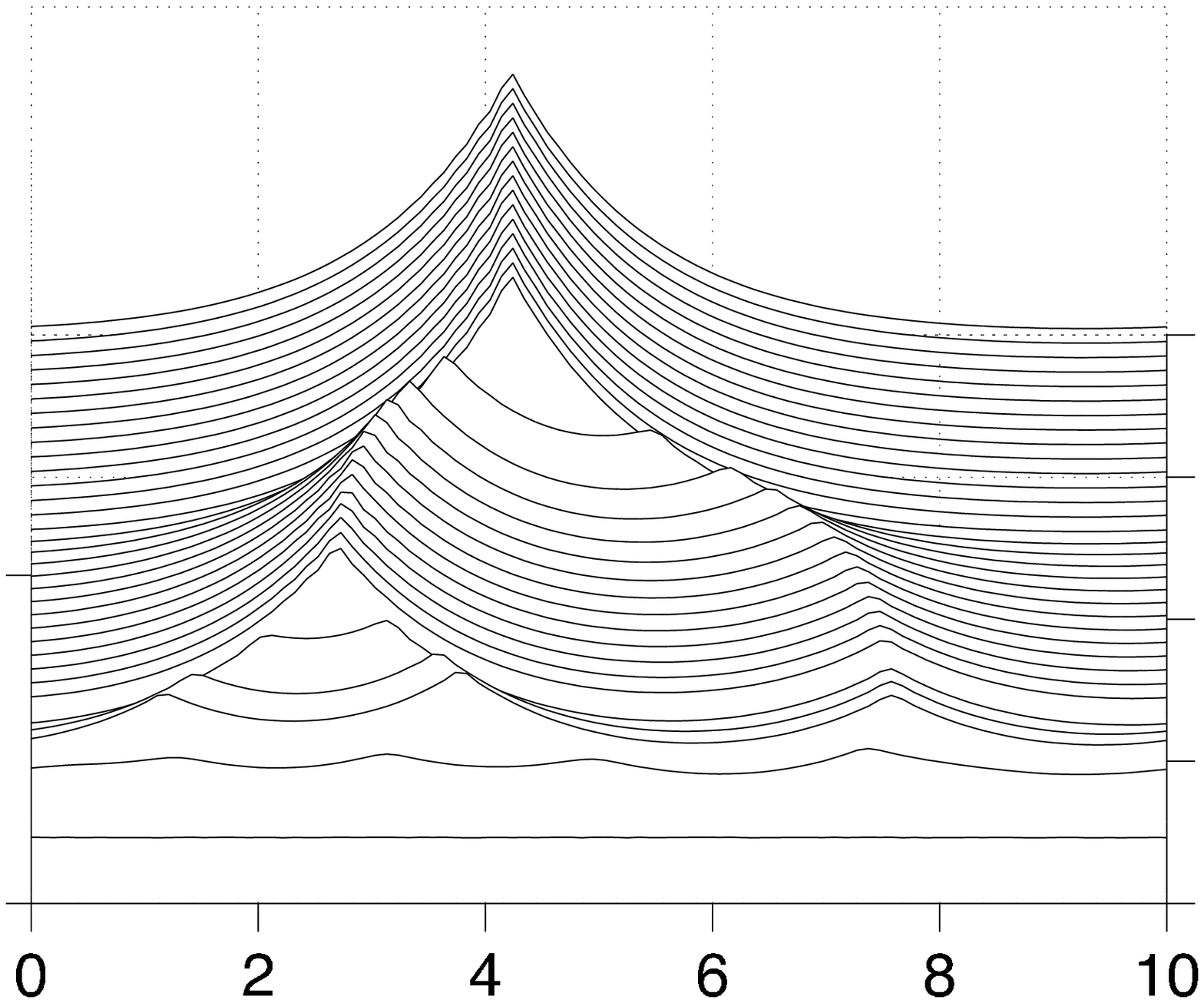}
\caption{\emph{Left:} A typical example of orienton formation in a  $d=1$ dimensional simulation. The color-code on the cylinder denotes local \emph{averaged} density $\overline{\rho}=H*\rho$ for a Helmholtz kernel $H=e^{-|x|}$: black is maximum density while white is $\overline{\rho}=0$.  Purple lines denote the three-dimensional vector $\overline{\bf M}= H* {\bf M}$. The formation of sharp peaks in averaged quantities corresponds to the formation of $\delta$-functions  in (\ref{rodrho}, \ref{DarcyGilbert}) according to (\ref{singansatz}).  Averaged quantities were chosen to avoid the necessity to represent $\delta$-functions. \emph{Right:} The corresponding waterfall plot for evolution of averaged density $\overline{\rho}=H*\rho$. 
Horizontal coordinate is space.  Sharp peaks correspond to the formation of $\delta$-function singularities in the density variable $\rho$.  }
\label{fig:orientons}
\end{center}
\end{figure}

We are now ready to write out explicit equations for the motion of magnets in $d=1$, $2$ or $3$ dimensions. 
In this case $G={\bf M}(x) \in \mathbb{R}^3$ -- average magnetization, $\rm ad_\mathbf{v} \mathbf{w}=\mathbf{v} \times \mathbf{w}$ and $\rm ad^*_\mathbf{v} \mathbf{w}=-\mathbf{v} \times \mathbf{w}$, and the
Lie algebra pairing is represented by the dot product of vectors in $\mathbb{R}^3$. Therefore, the equations for $(\rho, {\bf M})$ are obtained directly from (\ref{rhogen-simplified},\ref{Ggen-simplified}) and are written as follows: 
\begin{equation}
\frac{\partial \rho}{\partial t}
=
\text{\rm\large div}
\Bigg(
\rho\,
\bigg(
\mu_\rho\, \text{\large$\nabla$}\frac{\delta E}{\delta \rho}
+\,
\boldsymbol\mu_{\bf M}\cdot\text{\large$\nabla$}\frac{\delta E}{\delta \bf M}
\bigg)
\,
\Bigg)
\label{rodrho}
\end{equation}
and
\begin{equation}
\frac{\partial {\bf M}\,}{\partial t}
=
\text{\rm\large div}\Bigg({\bf M}\,\text{\large$\otimes$}
\left(\mu_\rho\text{\large$\nabla$}\frac{\delta E}{\delta \rho}
+
\boldsymbol\mu_{\bf M}
\cdot\text{\large$\nabla$}\frac{\delta E}{\delta {\bf M}}
\right)
 \Bigg)\!
+
{\bf M}\times
\boldsymbol{\mu}_{\bf M}\times\frac{\delta E}{\delta \bf M}
\label{DarcyGilbert}
\end{equation}
This is the generalization of the Landau-Lifshitz-Gilbert and Debye-H\"uckel equations we have sought. 
The main consequence of equations (\ref{DarcyGilbert})  is that they allow single particle solutions of the form
\begin{align}
\nonumber
\rho(\boldsymbol{x},t)&=w_\rho(t)\,\,\delta(\boldsymbol{x}-\boldsymbol{Q}
(t))\\
{\bf M}(\boldsymbol{x},t)&=\boldsymbol{w}_{\bf M}(t)\,\,\delta(\boldsymbol{x}-\boldsymbol{Q}
(t))
\label{singansatz}
\end{align}
where $w_\rho$\,, 
$\boldsymbol{w}_{\bf M}$ and $\boldsymbol{Q}
$ undergo the following
dynamics
\begin{align*}
\dot{w}_\rho&=0,
\quad\,\,\,
\dot{\boldsymbol{w}}_{\bf M}=\left(
\boldsymbol{w}_{\bf M}\times\boldsymbol\mu_{\bf M}\times\frac{\delta E}{\delta {\bf M}}\right)_{\boldsymbol{x}=\boldsymbol{Q}
}\!\!,
\quad\,\,\,
\dot{\boldsymbol{Q}}
=
-\left(\mu_\rho \text{\large$\nabla$}\frac{\delta E}{\delta \rho}
+
\boldsymbol\mu_{\bf M}\cdot\text{\large$\nabla$}\frac{\delta E}{\delta {\bf M}}\right)_{\boldsymbol{x}=\boldsymbol{Q}
}\!\!,
\end{align*}
These solutions have an important physical meaning, since they represent
particles that aggregate and align. This phenomenon is of fundamental importance
in the theory of anisotropic self-assembly.
There are special cases where these solutions emerge spontaneously from {\it
any} confined initial distribution. These behavior depends on the particular
choice of mobilities and energy functionals. A typical result of simulations is presented on Fig.~\ref{fig:orientons}. Starting with a random distribution in one dimension, we see the formation of sharp peaks in magnetization ${\bf M}$ (left figure)  and density $\rho$ (color code on the left and plot on the right), corresponding to the $\delta$-functions. 

\section{Moving dissipative oriented curves: new variables in geometric dynamics}
The structure of singular solutions allows an interesting geometric generalization, namely a possibility for singular solutions to be concentrated on a one-dimensional manifold (a curve) 
in the two- or three- dimensional space. This will give an interesting addition to the theory of exact geometric rods \cite{SiMaKr1988}, as our equations describe the \emph{dissipative} motion of a curve. Formally, we extend equations (\ref{singansatz}) and assume  that the solution is concentrated on a curve in space $\mathbf{r}=\mathbf{Q}(s,t)$, where $s$ is the arclength. Then, we are interested in the solutions of the form 
\begin{align}
\nonumber
\rho(\boldsymbol{x},t)&=w_\rho(s,t)\,\,\delta(\boldsymbol{x}-\boldsymbol{Q}
(s,t))\\
{\bf M}(\boldsymbol{x},t)&=\boldsymbol{w}_{\bf M}(s,t)\,\,\delta(\boldsymbol{x}-\boldsymbol{Q}
(s,t))
\label{singansatzcurve}
\end{align}
Multiplying each equation (\ref{rodrho}, \ref{DarcyGilbert}) by a test function and integrating produces the following equations of motion: 
\begin{eqnarray} 
\frac{\partial}{\partial t} w_\rho(s,t) & = & 0 
\label{rhoevolution}
\\
\frac{\partial}{\partial t} \boldsymbol{w}_{\bf M}(s,t)
&=&
 \boldsymbol{w}_{\bf M} \times 
\boldsymbol{\mu}_{\bf M}(\boldsymbol{Q}(s,t)) \times
\frac{\delta E}{\delta \bf M}(\boldsymbol{Q}(s,t)) 
\label{Mevolution}
\\
\frac{\partial}{\partial t} \boldsymbol{Q}(s,t)
&=&
\mu_\rho(s,t) 
\nabla \frac{\delta E}{\delta \rho} \Big(\boldsymbol{Q}(s,t) \Big) 
+\boldsymbol{\mu}_{\bf M}(s,t) \nabla \frac{\delta E}{\delta \bf M}(\boldsymbol{Q}(s,t))
\label{Qevolution0}
\end{eqnarray} 
In the simplest case of a binary energy functional,
\begin{equation} 
E=\frac{1}{2} \int \rho(\mathbf{x})\rho(\mathbf{x}') 
G_\rho \Big(\mathbf{x} -\mathbf{x}' \Big)
+
{\bf M}(\mathbf{x}){\bf M}(\mathbf{x}') G_{{\bf M}} \Big(\mathbf{x} -\mathbf{x}'\Big) d\,^3x\,\, d\,^3x'\,,
\label{binaryE}
\end{equation}
where $G_{\bf M}$ is assumed to be a scalar function, 
equation (\ref{Qevolution0}) assumes an especially simple and elegant shape -- see also \cite{HoPuSt2004}: 
\begin{eqnarray} 
\frac{\partial}{\partial t} \boldsymbol{Q}(s,t) 
&=&
\int 
\mu_\rho(s,t) \rho(s',t) 
\nabla G_\rho \Big(\boldsymbol{Q}(s,t), \boldsymbol{Q}(s',t) \Big) 
\nonumber
\\
& &\,+\
\boldsymbol{\mu}_{\bf M}(s,t) \cdot {\bf M} (s', t) 
\nabla 
G_{{\bf M}} \Big(\boldsymbol{Q}(s,t), \boldsymbol{Q}(s',t) \Big) 
\mbox{d} s' 
\, . 
\label{Qevolution} 
\end{eqnarray} 

\subsection{Numerical solution of curve evolution} 
To illustrate the concepts developed above, we perform a simulation of the spatio-temporal evolution for two oriented curves, having the simplest energy dependence 
given by (\ref{binaryE}).  In the future, we shall consider more sophisticated models of energy using the technique developed in (\ref{dEdmeq}) below. In computer simulations presented on Fig.~\ref{fig:twocurves}, we assumed centrally symmetric 
Lennard-Jones potential for the density interaction and the Helmholtzian $G_M( \bx)=\exp(- |\bx|)$ for the orientation part of energy. Initially, the curves are far away from each other and the ${\bf M}$ values on the curves are unrelated. Eventually, the curves collapse to two parallel lines and the ${\bf M}$ vectors on both curves align. 
\begin{figure}[htbp]
\begin{center}
\includegraphics[height=10cm]{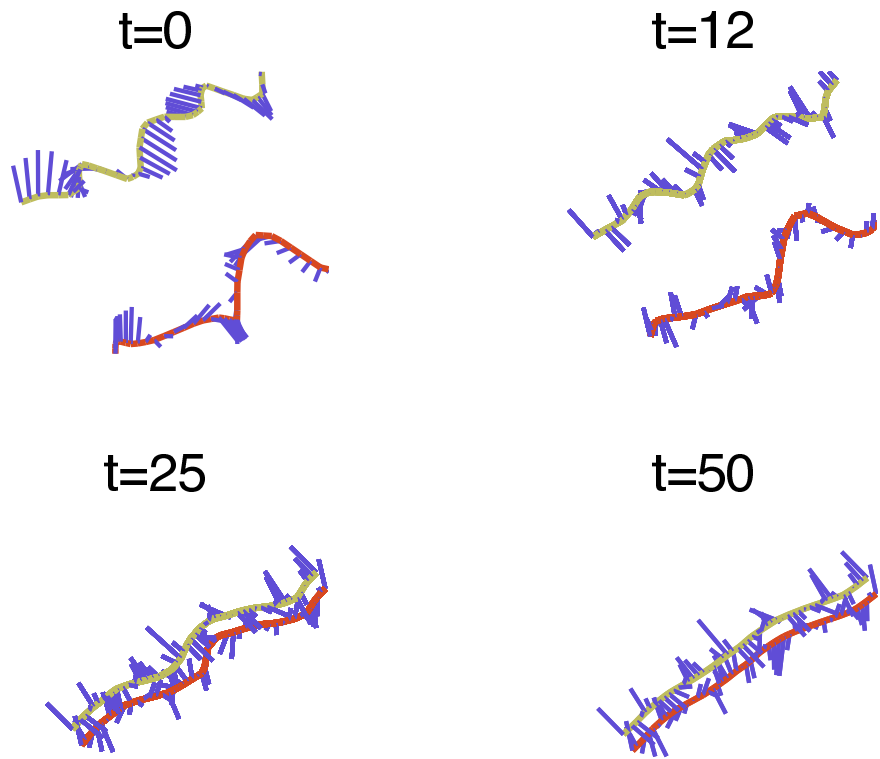}
\caption{An example of two oriented curves (red and green) attracting each other and unwinding at the same time.  The  blue vectors illustrate the vector ${\bf M} \in so(3)$ at each point on the curve. Time scale is arbitrary. }
\label{fig:twocurves}
\end{center}
\end{figure}

\section{Energy dependent on interactions of orientations}
Energy (\ref{binaryE}) provides insight, but it is too simple to describe many interesting physical phenomena, for example, dynamics of assembly of biologically relevant particles into curves like DNA. The energy of these particles depends on the position $\bx$ and orientation $\Lambda(\bx)$ of a frame  at a point $\bx$, and the \emph{relative} distances and  orientations  with respect to all other points $\bx'$.   We are thus compelled to deal with more interesting energies reflecting this physical fact, and this will lead us to some interesting mathematical consequences. The technical question here is how to compute $\delta E/\delta {\bf M}$, if the energy is given in terms of $\Lambda$. 
Of particular interest to us is the so-called anisotropic Lennard-Jones potential that has been developed in \cite{PaYa2005} and has been subsequently applied to produce self-organization of anisotropic particles into the shape of a double helix \cite{FeWa2007}. The idea of the method is the following. 

Consider two ellipsoids with the geometric centers at $\bx$ and $\bx '$ and orientation with respect to the fixed frame $\Lambda(\bx)$ and 
$\Lambda(\bx ')$, respectively. The method computes the distance $d\big(\bx,\bx ', \Lambda(\bx), \Lambda(\bx ')\big) $ between ellipsoidal  surfaces along the direction of $\bx - \bx'$. This is close to the minimal distance between the ellipsoids, but it is easier to compute. In principle, to make the depth of the potential well different for attracting and repulsive potential, one can consider two distances $d=d_1$ and $d=d_2$, corresponding to the attractive and repulsive parts of the ellipsoids. The distances $d_1$ and $d_2$ are defined as 
$d_i=\sigma_0 +|\bx-\bx'|+f_i(\Lambda(\bx,\lambda\bx'))$, where $\sigma_0$ is a given constant and $f_i$ are some (rather complex) functions. 
One then takes the potential 
to be the Lennard-Jones potential of the modified distances:  
\begin{equation} 
U\{ d\big(\bx,\bx ', \Lambda(\bx), \Lambda(\bx ')\big)\} 
=4 \epsilon_0 \Bigg[
 \Big( 
 \frac{\sigma_0}{d_1}
 \Big)^{12}
 - 
  \Big( 
 \frac{\sigma_0}{d_2}
 \Big)^6 
  \Bigg] 
\end{equation} 
An important feature of this potential that has not been noticed in previous works is its invariance with respect to $SO(3)$ rotations about each point on the curve. Thus, the energy functional may be written as 
\begin{equation} 
E= 
\frac{1}{2} 
\int 
\rho(\bx) 
\rho(\bx') 
U \Big( \bx,\bx ', \Lambda(\bx), \Lambda(\bx ') \Big) 
\mbox{d} \bx 
\mbox{d} \bx' 
\, , 
\end{equation} 
where $U$ is some scalar function of its arguments. This energy is still \emph{binary}, {\emph i.e.} it is a direct sum of pairwise interaction of all particles, which is known to be an rather crude approximation for real systems. However, we believe it is important to first understand this $SO(3)$-invariant energy. Later, the equations of motion for more general energies may be obtained by direct generalization. 
Invariance with respect to the rotation group requires that the energy of particle interaction assumes the following form: 
\begin{equation} 
E= 
\frac{1}{2} 
\int 
\rho(\bx) 
\rho(\bx') 
U\Big(  |\bx-\bx ' |, \xi(\bx, \bx')  \Big)
\mbox{d} \bx 
\mbox{d} \bx' 
\, , 
\label{EnergyLJ} 
\end{equation} 
where we have defined 
\begin{equation} 
\xi(\bx,\bx')=  \Lambda(\bx)^{-1} \Lambda(\bx ') \in SO(3)  
\, . 
\label{xidef}
\end{equation} 
\begin{remark} 
Note that we have assumed that the energy does not depend directly on a Lie algebra element 
$\widehat{\bf M}= \Lambda^{-1} \nabla \Lambda$. That means, in particular, that there is no elastic energy caused by particle motion. The introduction of  elastic energy complicates the equations and will be considered in future work. 
\end{remark} 
Now, we need to compute $\delta E/\delta \Lambda$. We proceed as follows. Define $\Sigma(\bx) = \Lambda^{-1}(\bx) \delta \Lambda(\bx) \in so(3)$. Then, 
notice that 
\begin{align} 
\delta \xi & = \delta \left( \Lambda^{-1}(\bx) \Lambda(\bx') \right)
\nonumber 
\\
 &=  
 -  \Lambda^{-1}(\bx) \delta \Lambda(\bx)  \Lambda^{-1}(\bx) \Lambda(\bx') 
 + \Lambda^{-1} (\bx) \delta \Lambda(\bx') 
 \nonumber 
 \\
 & = 
- \Sigma(\bx)   \xi(\bx,\bx') +\Lambda^{-1}(\bx) \Lambda(\bx') \Lambda^{-1}(\bx') \delta \Lambda(\bx') 
\nonumber 
\\ 
& = 
 - \Sigma(\bx)   \xi(\bx,\bx') + \xi(\bx,\bx') \Sigma(\bx')
 \, , 
\end{align} 
since $\xi =  \Lambda^{-1}(\bx) \Lambda(\bx') \in SO(3)$.
 Thus, 
\begin{align} 
\delta E=&\,
\int
\rho(\bx)
\rho(\bx') 
{\rm tr} \left( \left(  \frac{\partial  U}{\partial \xi} (\bx,\bx') \right)^T 
\Big\{ 
- \Sigma(\bx)   \xi(\bx,\bx') + \xi(\bx,\bx') \Sigma(\bx')
\Big\} 
\right)
\mbox{d} \bx  
\mbox{d} \bx'
\nonumber
\\
=&\,
\int
\rho(\bx)
\rho(\bx') 
{\rm tr} 
\left( 
-\left( 
\frac{\partial  U}{\partial \xi} (\bx,\bx') \xi (\bx,\bx')^T 
\right)^T \Sigma(\bx) 
\right)
\mbox{d} \bx 
\mbox{d} \bx' 
\nonumber
\\
&
+
\int
\rho(\bx)
\rho(\bx') 
{\rm tr} 
\left( 
 \left(  \xi(\bx,\bx')^T \frac{\partial  U}{\partial \xi}(\bx,\bx') \right)^T
 \Sigma(\bx') 
\right) 
\mbox{d} \bx 
\mbox{d} \bx' 
\, . 
\end{align}
In the last term, $\bx $ and $\bx'$ were exchanged to make the variation $\Sigma(\bx')$ a function of $\bx$. 
When making this exchange, we must remember that 
\[ 
\xi(\bx,\bx')=\Lambda(\bx)^{-1} \Lambda(\bx')=\xi^{-1} (\bx',\bx)= \xi^T(\bx',\bx) \, , 
\] 
and similarly, 
\[ 
 \frac{\partial  U}{\partial \xi}(\bx,\bx')=  \left( \frac{\partial  U}{\partial \xi}(\bx',\bx) \right)^T 
 \,.
\] 
Consequently, we find, 
\begin{align*}
\delta E=& 
\int
\rho(\bx)
\rho(\bx') 
{\rm tr} 
\left( 
-\left( 
\frac{\partial  U}{\partial \xi} (\bx,\bx') \xi(\bx,\bx')^T 
\right)^T 
\Sigma(\bx) 
\right) 
\\
&+
\int
\rho(\bx)
\rho(\bx') 
{\rm tr} 
\left( 
 \left( \xi(\bx,\bx') \left( \frac{\partial  U}{\partial \xi}(\bx,\bx') \right)^T  \right)^T 
\Sigma(\bx) 
\right) 
\mbox{d} \bx 
\mbox{d} \bx' 
\, . 
\end{align*}
Thus, collecting terms leads to
\[ 
\delta E= 
\int
\rho(\bx)
\rho(\bx')
\left< 
-\frac{\partial  U}{\partial \xi} (\bx,\bx') \xi(\bx,\bx')^T 
+
 \xi(\bx,\bx') \left( \frac{\partial  U}{\partial \xi}(\bx,\bx') \right)^T 
, 
\, 
\Sigma (\bx) 
\right> 
\mbox{d} \bx 
\mbox{d} \bx' 
\, . 
\] 
If we now turn from the variations in $\Sigma$ to variations in $\Lambda$, we find $\Lambda$ in front of the integral, and the final formula for the variation of energy is thus 
\begin{equation} 
\frac{\delta E}{\delta \Lambda}  = 
\Lambda(\bx)  \rho(\bx) \int 
\rho(\bx') 
 \bigg\{ 
-\frac{\partial  U}{\partial \xi}(\bx,\bx') \xi^T(\bx,\bx') 
+\xi(\bx,\bx') \left( \frac{\partial  U}{\partial \xi}(\bx,\bx') \right)^T 
\bigg\} 
\mbox{d} \bx' 
\, . 
\label{dEdlambda}
\end{equation} 

We have computed the variations with respect to $\Lambda$. We are not finished yet, as 
 the equations of motion of a string are formulated on the Lie algebra, and not the tangent manifold to Lie group. Thus, we need to connect 
$\delta E/\delta \Lambda$ to $\delta E / \delta \widehat{{\bf M}}$.%
\footnote{Here, the hat denotes the usual isomorphism between vectors in $\mathbb{R}^3$ and elements of Lie Algebra $so(3)$, so $\widehat{{\bf M}} \in so(3)$ and $ {\bf M} \in \mathbb{R}^3$.} To do that, we use the definition of the left-invariant tangent space at the identity,
\begin{equation} 
\widehat{{\bf M}}_i
=\Lambda^{-1} \frac{\partial\Lambda}{\partial x^i} 
\in so(3)
\, , \quad
i=1,2,\dots,d
\,,
\label{Mi-puregauge}
\end{equation} 
where $x^i$ is the $i$-th coordinate in space. The Lie algebra element is then a `vector' of $d$ antisymmetric matrices, where $d$ is the dimension of the space. We have, by the usual change of variables (summation over repeated indices is assumed), 
\begin{equation} 
\delta E= \left< \frac{\delta E}{\delta \Lambda} \, , \, \delta \Lambda \right>  
= 
\left< \frac{\delta E}{  \delta \widehat{{\bf M}}_i} \, , \, \delta \widehat{{\bf M}}_i  \right>  
 \, , 
\label{changevar} 
\end{equation} 
where $\delta E / \delta \widehat{{\bf M}}\in so(3)^*$.
Denote again  $\Sigma(\bx) = \Lambda^{-1} (\bx) \delta \Lambda(\bx) \in so(3)$. Then, 
\[ 
 \delta \widehat{{\bf M}}_i =- \Sigma \widehat{{\bf M}}_i + \Lambda^{-1} \frac{\partial  }{\partial x^i} 
 \delta \Lambda
 \, . 
\] 
Thus, (\ref{changevar}) becomes 
\begin{align} 
\left< \frac{\delta E}{\delta \Lambda} \, , \, \delta \Lambda \right> 
 = &
\left< \frac{\delta E}{  \delta \widehat{{\bf M}}_i} \, , \,  
- \Sigma \widehat{{\bf M}}_i+ \Lambda^{-1} \frac{\partial  }{\partial x^i}  \Lambda \Sigma  \right> 
\nonumber 
\\ 
&= 
\left< \frac{\delta E}{  \delta \widehat{{\bf M}}_i} \, , \, \left[ \widehat{{\bf M}}_i \, , \, \Sigma \right]  \right> 
- 
\left< \frac{\partial}{\partial x^i} \frac{\delta E}{  \delta \widehat{{\bf M}}_i} \, ,  \, \Sigma \right> \, . 
\label{dedlambdacalc} 
\end{align} 
Remembering that $\Sigma =\Lambda^{-1} \delta \Lambda$ and setting $\Lambda$ using the pairing properties \\
$<a,\Lambda b>= <\Lambda^T a,b>$ leads to the following expression (as $(\Lambda^{-1})^T=\Lambda$): 
\begin{equation} 
\frac{\delta E}{\delta \Lambda} = 
\Lambda  \left( {\rm ad}^*_{\widehat{{\bf M} }_i }\frac{\delta E}{  \delta \widehat{{\bf M}}_i} - \frac{\partial}{\partial x^i} \frac{\delta E}{  \delta \widehat{{\bf M}}_i} 
\right).
\label{dEdlambda0} 
\end{equation} 
This can be written in a familiar way in terms of the vector product in $\mathbb{R}^3$ as 
\begin{equation} 
\frac{\delta E}{\delta \Lambda} =  
- 
\Lambda \left( 
{\bf M}_i \times \frac{\delta E}{  \delta {\bf M}_i} + \frac{\partial}{\partial x^i} \frac{\delta E}{  \delta {\bf M}_i} 
\right)^{\!\bf \widehat{\,\,\,}  } 
\, , 
\label{dEdlambda1} 
\end{equation} 
which is an extension of the formula (6.18d) of \cite{SiMaKr1988}. 

Combining equations (\ref{dEdlambda1}) and (\ref{dEdlambda}) yields an auxiliary equation for $\delta E/ \delta {\bf M}$ that is valid for any value of $\bx$:
\begin{align} 
\left( 
{\bf M}_i (\bx)  \times \frac{\delta E}{  \delta {\bf M}_i} (\bx) + \frac{\partial}{\partial x^i} \frac{\delta E}{  \delta {\bf M}_i} (\bx) 
\right)^{\!\bf \widehat{\,\,\,}}
= &-
\rho(\bx) \int \rho(\bx') 
\Big\{ 
-\frac{\partial  U}{\partial \xi}(\bx, \bx')  \xi(\bx, \bx')^T
\nonumber 
\\
&+\xi(\bx, \bx') \left( \frac{\partial  U}{\partial \xi}(\bx, \bx') \right)^T 
\Big\} 
\mbox{d} \bx' 
\, . 
\label{dEdmeq} 
\end{align} 
where, again,  $\xi(\bx, \bx'):=\Lambda^{-1}(\bx) \Lambda(\bx') \in SO(3)$. The left hand side of this equation is the $so(3)^*$-valued covariant divergence of the $so(3)^*$ vector $\delta E/ \delta {\bf M}_i$, which is also a 3-by-3 antisymmetric matrix for each $i=1,2,\dots,d$. Likewise, the expression in curly brackets is an antisymmetric 3-by-3 matrix, as one may verify by taking the transpose. Thus equation (\ref{dEdmeq}) is consistent. In fact, equation (\ref{dEdmeq}) is reminiscent of the Yang-Mills Gauss Law for an $so(3)^*$-valued gauge charge. A similar equation appears in the gauge-theoretical formulation of the dynamical equations  for a spin-glass \cite{HoKu1988}. The gauge freedom remaining in equation  (\ref{dEdmeq}) is removed by noticing that ${\bf M}_i$ defined in (\ref{Mi-puregauge}) is a pure gauge field. Consequently, the connection associated with the $so(3)$-valued covariant divergence in (\ref{dEdmeq}) must have no curvature. Hence, for any $i,j,$ we find the following consistency conditions
\begin{equation} 
\frac{\partial \widehat{ {\bf M}}_i}{\partial x^j} - 
\frac{\partial \widehat{{\bf M}_j}}{\partial x^i} 
= 
\left[ 
\widehat{{\bf M}}_i \, , \, \widehat{{\bf M}}_j 
\right] = \widehat{{\bf M}}_i  \widehat{{\bf M}}_j - \widehat{{\bf M}}_j  \widehat{{\bf M}}_i
\label{consistency} 
\, , 
\end{equation} 
which close the system (\ref{dEdmeq}).

\begin{remark}
It is interesting to note the similarity between the variable 
$\xi=\Lambda^{-1}(\bx) \Lambda(\bx')$ and  its time analogue  $\Lambda(t)^{-1} \Lambda(t+ \Delta) $, which is commonly used in discrete-time Moser-Veselov integrators \cite{MoVe1991} and their extensions  \cite{Di2007}. Further study of  the geometric nature of (\ref{dEdmeq}) appears intriguing and will be undertaken in the future. 
\end{remark}

Let us see how (\ref{dEdmeq}) reduces on the singular solutions (\ref{singansatzcurve}). 
Assuming that the potential $U$ is such that $\partial U/\partial \xi$ remains finite on the singular solutions, we substitute (\ref{singansatzcurve}) into  (\ref{dEdlambda}) to find 
\begin{align} 
\frac{\delta E}{\delta \Lambda} (\bx) & = \Lambda(\bx) 
w_\rho(s,t) \delta \big(\bx-\boldsymbol{Q}(s,t) \big)  
\nonumber 
\\ 
& \int 
w_\rho(s',t)  
 \Big\{ 
-\frac{\partial  U}{\partial \xi} \big( s, s' \big) \xi^T \big( s, s' \big) 
+\xi (s, s') \left( \frac{\partial  U}{\partial \xi} (s,s') \right)^T 
\Big\} 
\mbox{d} \bx' 
\, , 
\label{dEdlambdacurve}
\end{align} 
where we have denoted 
\[ 
\xi(s,s'):= \xi \big( \boldsymbol{Q}(s,t) , \boldsymbol{Q}(s' ,t) \big) 
\quad \mbox{and} 
\quad 
\frac{\partial  U}{\partial \xi} ( s, s' ) 
:= 
\frac{\partial  U}{\partial \xi}  \big( \boldsymbol{Q}(s,t) , \boldsymbol{Q}(s' ,t) \big) 
\, . 
\] 
Upon multiplying each side of the equation (\ref{dEdmeq}) by an arbitrary function $\phi(\bx)$, 
integrating over the whole space and equating the coefficients of $\phi\big( \boldsymbol{Q}(s,t)\big)$, we obtain the following relation: 
\begin{align} 
\left( 
\boldsymbol{w}_{\bf M} (s)  \times \frac{\delta E}{  \delta {\bf M}} (s) + \frac{\partial}{\partial s} \frac{\delta E}{  \delta {\bf M} }(s) 
\right)^{\!\bf \widehat{\,\,\,}}
= &-
w_\rho(s) \int w_\rho(s') 
\Big\{ 
-\frac{\partial  U}{\partial \xi}(s,s')  \xi(s,s')^T
\nonumber 
\\
&+\xi(s,s') \left( \frac{\partial  U}{\partial \xi}(s,s') \right)^T 
\Big\}  \mbox{d} s'
\, . 
\label{dEdmeqcurve} 
\end{align} 
This is the equation for the computation of the variations $\delta E/\delta {\bf M}$ on the curve that we have sought. The analogue for curves of the consistency condition (\ref{consistency}) then closes the system. Future work will explore its dynamics and numerical solutions.

\section{Conclusions}
In this paper, we explained the derivation of continuum equations for dissipation using the moment approach in kinetic (Vlasov) equations. We showed how to introduce dissipation in Vlasov equations based on a  geometric generalization 
of the Darcy's law of motion (force being proportional to velocity). We derived the continuum equations of motions in general settings and then formulated those equations for the particular case of self-organization of magnetic particles. We showed the existence and spontaneous emergence of generalized solutions concentrated on delta-functions. We also showed that the equations possess singular solutions concentrated on one-dimensional manifolds (curves) and derived equations of motion for these curves. 

Of particular interest for future work are two questions. The first question is, Can solutions defining singular curves appear from random initial conditions in two or three dimensions? The positive answer to this question would demonstrate the possibility of controlling magnetic self-assembly of dispersed particles into ``magnetic strings", which would be very interesting technologically. 

The second question is, Can we consistently incorporate both inertia and dissipative terms in the theory of exact geometric rods? Hopefully, this can be accomplished by starting with the dissipative Vlasov equation containing the inertia terms, repeating the 
procedure of \cite{GiHoKu1982, GiHoKu1983} keeping the momentum term and closing the system using either truncation or cold plasma approximation. An interesting task would be to compare the resulting dissipative equations with those of the exact geometric rods in the absence of dissipation \cite{SiMaKr1988}. 

\subsection*{Acknowledgements} 
This work was presented as a contribution to the conference celebrating 60th year birthday of D. D. Holm. We would like to thank the scientific committee (R. Camassa, M. Hyman, J. Marsden, T. Ratiu and E. Titi) for their excellent work in organizing this conference. We are also grateful to  the Center Interfacultaire Bernoulli for providing financial support of the conference, and to the staff of the Center for running the events of the conference so cheerfully and efficiently. 
The authors were partially supported by NSF grant NSF-DMS-05377891. The work of 
DDH was also partially supported  by the US Department of Energy, Office of Science, Applied Mathematical Research, and the Royal Society of London Wolfson Research Merit Award. VP acknowledges the support of A. v. Humboldt foundation, the hospitality of Department for Theoretical Physics, University of Cologne, and the European Science Foundation for partial support through the MISGAM program. 

\bibliographystyle{unsrt}

\begin{thebibliography}{99}

\bibitem{Sunetal2000}  S. Sun, C.B. Murray, D. Weller, L. Folks, and A. Moser, Monodisperse FePt Nanoparticles and Ferromagnetic FePt Nanocrystal Superlattices, 
\emph{Science}, {\bf 287}, p. 1989-1992 (2000). 

 \bibitem{Zengetal2002} H. Zeng, J. Li, J. P. Liu, Z. L. Wang and S. Sun,  Exchange-coupled nanocomposite magnets by nanoparticle self-assembly, 
\emph{Nature}  {\bf 420}, 395 - 398 (2002).

\bibitem{BaCoKe2005} J. V. Barth G. Costantini and  K. Kern, 
Engineering atomic and molecular nanostructures at surfaces
\emph{Nature} {\bf 437}, 671 - 679 (2005).

\bibitem{Ba2000} G. K. Batchelor, An Introduction to Fluid Dynamics, 
\emph{Cambridge University Press}, (2000). 

\bibitem{Darcy1856}
Henry P. G. Darcy is best known for his empirical law on fluid flow through porous media, published as an appendix to his book {\it Les Fontaines publiques de la ville de Dijon}. Darcy had built a water supply system for Dijon in 1840, and in 1856, shortly before his death, he wrote the book to guide other engineers in constructing similar projects. Darcy's book was recently translated into English by P. Bobeck [2003] {\it Henry Darcy and the Public Fountains of the City of Dijon}, Geotechnical Translations, Austin, Texas.


\bibitem{DeHu1923}  P. Debye, E. H\"uckel, The Interionic Attraction Theory of Deviations from Ideal Behavior in Solution II,
\emph{Zeitschrigt f\"ur Physik}, pp. 25-97 (1924).

\bibitem{Ot2001} 
F. Otto,  The geometry of dissipative evolution equations: the porous medium equation. Comm. Partial Differential Equations 26 (2001) 
101-174

\bibitem{HoPu2005} D.~D.~Holm and V.~Putkaradze, 
Aggregation of finite-size particles with variable mobility,
Phys Rev Lett, 95 (2005) 226-106. 

\bibitem{HoPu2006} D.~D.~Holm and V.~Putkaradze, Formation of clumps and patches in self-aggregation of finite size particles,
Physica D,   220  (2006) 183-196. 

\bibitem{Gi2004} T. L. Gilbert, A Phenomenological Theory of Damping in Ferromagnetic Materials, IEEE Transactions on Magnetics, {\bf .40}, (6), 2004. 
\bibitem{LaLi1935} 
L. D.Landau and E. M.Lifshitz, ÒOn the theory of the dispersion of magnetic permeability in ferromagnetic bodies,Ó Phys. Z. Sowjet., {\bf 8},pp.153Ð169,(1935); see also L. D. Landau, Collected Papers., \emph{ed.by D. terHaar}.Gordon and Breach, New York, 1967, 101.
\bibitem{Be1973}
D. J. Benney, \textit{Properties of long nonlinear waves}.
Stud. App. Math. 52 (1973) 45. 
\bibitem{Bl1950}
N. Bloembergen,
On the ferromagnetic resonance in
nickel and Supermalloy,  \emph{Phys. Rev.}  {\bf 78}, 572 (1950).


\bibitem{CoOlTo1955}
R. S. Codrington, J. D. Olds, and H. C. Torrey, "Paramagnetic resonance in organic free radicals at low fields," \emph{Phys. Rev.} {\bf  95}, 607 (1954). 

\bibitem{Chavanis} P.~H.~ Chavanis, P. Lauren{\c c}ot, M. Lemou,
 {\it Chapman–-Enskog derivation of the generalized Smoluchowski equation}.
 Physica A 341 (2004) 145-–164 

\bibitem{HoPuTr2007-CR} 
D.~D.~Holm and V.~Putkaradze, C.~Tronci, {\it
Geometric dissipation in kinetic equations.} C. R. Acad. Sci. Paris, to appear
(2007) (also at arXiv:0705.0765)

\bibitem{Ka1991}
H. E. Kandrup,  The secular instability of axisymmetric collisionless star cluster. \emph{Astrophys. J.} {\bf 380}  pp. 511-514 (1991).

\bibitem{BlBrRa1992} A. M. Bloch, R. W. Brockett and T.S.Ratiu, 
{\em Comm. Math. Phys. }  {\bf 147}, 57-74 (1992). 


\bibitem{BlKrMaRa1996} A. Bloch, P. S. Krishnaprasad, J. E. Marsden, T. S. Ratiu,  The Euler-PoincarŽ equations and double bracket dissipation,
\emph{Comm. Math. Phys.} {\bf 175} pp.1-42 (1996).


\bibitem{Ka1984} A. N. Kaufman,   Dissipative Hamiltonian systems: a unifying principle,  \emph{ Phys. Lett. A},  {\bf 100}  pp. 419-422 (1984).
\bibitem{Mo1984} P.~J.~Morrison, Bracket formulation for irreversible classical fields,  \emph{ Phys. Lett. A},  {\bf 100}  pp. 423-427 (1984).

\bibitem{Gr1984}
M. Grmela, Bracket formulation of dissipative fluid mechanics equations, {\it Phys. Lett. A} {\bf 102}, 355-358 (1984). 

\bibitem{Gr1993a}
M. Grmela,  
Weakly nonlocal hydrodynamics, {\it Phys. Rev. E} {\bf 47}, 351-365 (1993). 

\bibitem{Gr1993b}
M. Grmela,  Thermodynamics of driven systems.  {\it Phys. Rev. E} {\bf 48}, 919-930 (1993). 

\bibitem{Ku1987}
B. A. Kupershmidt,  {\it Hydrodynamical Poisson brackets and local Lie algebras.}
\emph{Phys. Lett. A}  {\bf 121}, no. 4, 167--174 (1987).

\bibitem{GiHoTr2005}
J. Gibbons, D. D.  Holm and C.  Tronci, {\it Singular solutions for geodesic flows of Vlasov moments}, Proceedings of the MSRI workshop \textit{Probability, geometry and integrable systems}, in celebration of Henry McKean's $75^\text{th}$ birthday, Cambridge University Press, Cambridge, 2007 (in press, also at arXiv:nlin/0603060)

\bibitem{GiHoTr07}
Gibbons, J.; Holm, D. D.; Tronci, C. {\it Vlasov moments, integrable systems
and singular solutions}, Phys. Lett. A 2007 (in press, also at arXiv.org:0705.3603)

\bibitem{VaCaYo1989} 
G. K. Vallis, G. F.  Carnevale,  and W. R. Young,   Extremal energy properties and construction of stable solutions of the Euler equations. {\it J. Fluid Mech.} {\bf 207}, 133-152 (1989).  

\bibitem{MaMo1980}
W. H. Matthaeus,  and D. Montgomery, 
Selective decay hypothesis at high mechanical and magnetic Reynolds numbers,
{\it Ann. N.Y. Acad. Sci.} {\bf 357} 203-222 (1980).

\bibitem{HoMaRaWe1985} 
D. D. Holm, J. E.  Marsden,  T. S. Ratiu, and A. Weinstein, Nonlinear stability of fluid and plasma equilibria. 
{\it Phys. Rep.} {\bf 123}, 1-116 (1985). 


\bibitem{Kr1967}
R. H. Kraichnan, 
Inertial ranges in two-dimensional turbulence,
{\it Phys. Fluids} {\bf 10}, 1417-1423 (1967).






\bibitem{Ta1986}
 J. B. Taylor,  
Relaxation and magnetic reconnection in plasmas, 
{\it Rev. Mod. Phys.} {\bf 58}, 741 - 763 (1986). 

\bibitem{HoMcRo1985} 
 B. J. Hoskins, M. E. McIntyre and A. W.  Robertson, 
On the use and significance of isentropic potential vorticity maps,
{\it Quart. J. Roy. Met. Soc.} {\bf 111}, 877-946. (1985). 

\bibitem{Yo1987}
W. R. Young,  
Selective decay of enstrophy and the excitation of barotropic waves in a channel, {\it J. Atmos. Sci.} {\bf 44}, 2804Ð2812.

\bibitem{HoPu2007} D.~D.~Holm and V.~Putkaradze, 
Interaction of particles with noncentral potential: gradient flows and singular solutions for evolution of geometric continuum quantities, Physica D, to appear (2007) (also at arXiv:nlin/0608054).  




\bibitem{TuSl2007} V. Tuberkevich and A. Slavin, Nonlinear phenomenological model of magnetic dissipation for large precession angles: Generalization of the Gilbert model, \emph{Phys. Rev. B}, {\bf 75}, 014440 (2007). 


\bibitem{GiHoKu1982}
Gibbons, J.; Holm, D. D.; Kupershmidt, B. A. \textit{Gauge-invariant
Poisson brackets for chromohydrodynamics.} Phys. Lett. A {\bf 90}  no. 6,
281--283 (1982).

\bibitem{GiHoKu1983} 
J. Gibbons,  D. D. Holm, and  B. A. Kupershmidt,  \textit{The Hamiltonian
structure of classical chromohydrodynamics.} Phys. D {\bf 6} (1982/83), no. 2,
179--194.

\bibitem{HoPuTr2007} 
D.~D.~Holm and V.~Putkaradze, C.~Tronci, {\it
Double bracket dissipation in kinetic theory for particles with anisotropic interactions.} Proceedings of the Summer School and Conference on Poisson Geometry, 4-22 July 2005, Trieste, Italy (also at arXiv:0707.4204)



\bibitem{SiMaKr1988}
   J.~C.~Simo,  J.~E.~Marsden  and P.~S.~Krishnaprasad
  The Hamiltonian structure of nonlinear elasticity: The material and convective representations of solids, rods, and plates,
\emph{Arch. Rat. Mech. Anal}
   {\bf 104},
  pp.125--183,
   (1988).

\bibitem{HoPuSt2004} D. D. Holm, V. Putkaradze and S. Stechmann, Rotating concentric circular peakons, \emph{Nonlinearity} {\bf 17} 2163-2186 (2004). 

\bibitem{PaYa2005} L.~Paramonov and S.~N.~Yaliraki, The directional contact distance of two ellipsoids: coarse-grained potentials for anisotropic interactions, \emph{J. Chem. Pys.}, {\bf 123}, 194111 (2005). 

\bibitem{FeWa2007} S. N. Fejer and D.~J.~Wales, Helix Self-assembly from Anisotropic Particles, \emph{Phys. Rev. Lett}, {\bf 99}, 086106 (2007). 

\bibitem{HoKu1988} 
D. D. Holm and B. Kupershmidt,
The analogy between spin glasses 
and Yang-Mills fluids, 
{\it J. Math Phys.} {\bf 29} (1988) 21--30.

\bibitem{MoVe1991}
J. Moser and A. Veselov,  Discrete version of some classical integrable systems and factorization of matrix polynomials. 
{\it Commun. Math. Phys.} {\bf 139} (1991), 217-243.

\bibitem{Di2007} M. Dixon, Geometric Integrators for Continuum Mechanics, \emph{Ph. D Thesis}, Imperial College, London (2007). 
\rem{
\bibitem{BlBrGlRaRy03}
Blaskiewicz, M.; Brown, K.; Glenn, J. W.; Raka, E.; Ryan, J. {\it Spill structure in intense beams.} Proceedings of the Particle Accelerator Conference Vol.
4 (2003), 2595--2597. 

\bibitem{BlIs}
Bloch, A. M.; Iserles, A. {\it On an isospectral Lie-Poisson system and its Lie algebra.} Found. Comput. Math. 6 (2006), no. 1, 121--144.

\bibitem{BlIsMaRa05}
Bloch, A. M.; Iserles, A.; Marsden, J.E.; Ratiu T. S.
\textit{A class of integrable geodesic flows on the symplectic group and the symmetric matrices.} arXiv.org:math-ph/0512093 (2005).

\bibitem{CaHo1993}
Camassa, R.; Holm, D. D. \textit{An integrable shallow water equation
with peaked solitons.} Phys. Rev. Lett. 71 (1993), no. 11, 1661--1664.

\bibitem{CaMaPu2002}
Caprino, S.; Marchioro, C.; Pulvirenti, M.
{\it On the two-dimensional Vlasov-Helmholtz equation with infinite mass.} Comm. Partial Differential Equations 27 (2002), no. 3-4, 791--808.

\bibitem{CeHoHoMa1998}
Cendra, H.; Holm, D. D.; Hoyle, M. J. W.; Marsden, J. E. \textit{The
Maxwell-Vlasov equations in Euler-Poincar\'e form.} J. Math. Phys. 39
(1998), no. 6, 3138--3157.

\bibitem{CeMaPeRa}
Cendra, H.; Marsden, J. E.; Pekarsky, S.; Ratiu, T. S.
\textit{Variational principles for Lie-Poisson and Hamilton-Poincar\'e
equations.} Mosc. Math. J. 3 (2003), no. 3, 833--867.

\bibitem{ChLiZh2005} Chen, M.; Liu, S.; Zhang Y. 
{ \it A two-component generalization of the Camassa-Holm equation and its solutions.}
Lett. Math. Phys. 75 (2006), no.1, 1-15.  ArXiv:nlin.SI/05010528

\bibitem{CoDaHoMa04}
Cousineau, S.; Danilov, V.; Holmes, J.; Macek, R. {\it Space-charge-sustained
microbunch structure in the Los Alamos Proton Storage Ring.} Phys. Rev. ST Accel. Beams 7 (2004), no. 9, 094201.

\bibitem{Dragt} 
Dragt, A. J.; Neri, F.; Rangarajan, G.; Douglas, D. R.; Healy, L. M.;
Ryne, R. D. \textit{Lie algebraic treatment of linear and nonlinear beam
dynamics.} Ann. Rev. Nucl. Part. Sci. 38 (1990), no. 38, 455--496.

\bibitem{DuON1999} 
Dubin, D. H. E.; O'Neil, T. M.
\textit{Trapped nonneutral plasmas, liquids, and crystals (the thermal
equilibrium states)}.
Rev. Mod. Phys., 71 (1990), no. 1, 87-172.

\bibitem{Falqui06} G. Falqui, 
On a Camassa-Holm type equation with two dependent variables, 
{\em J. Phys. A: Math. Gen.} {\bf 39} 
(2006), 327Ð342, nlin.SI/0505059. 

\bibitem{Gi1981}
Gibbons, J. \textit{Collisionless Boltzmann equations and integrable
moment equations.} Phys. D 3 (1981), no. 3, 503--511.


\bibitem{GiHoTr05}
Gibbons, J.; Holm, D. D.; Tronci, C. {\it Singular solutions for geodesic flows of Vlasov moments}, Proceedings of the MSRI workshop \textit{Probability, geometry and integrable systems}, in celebration of Henry McKean's $75^\text{th}$ birthday, Cambridge University Press, Cambridge, 2007 (in press, also at arXiv.org:nlin/0603060)


\bibitem{GiTs1996}
Gibbons, John; Tsarev, Serguei P. {\it Reductions of the Benney equations.} Phys. Lett. A 211 (1996), no. 1, 19--24

\bibitem{HoMa2004} 
Holm, D. D.; Marsden, J. E. 
\textit{Momentum maps and measure valued solutions (peakons, filaments,
and sheets) of the Euler-Poincar«e equations
for the diffeomorphism group.}
In {\it The Breadth of Symplectic and Poisson Geometry, A Festshrift for
Alan Weinstein}, 203-235, { Progr. Math.}, {232}, J.E. Marsden
and T.S. Ratiu, Editors, Birkh\"auser Boston, Boston, MA, 2004. 

\bibitem{HoMaRa}
Holm, D. D.; Marsden, J. E.; Ratiu, T. S. \textit{The Euler-Poincar\'e
equations and semidirect products with applications to continuum
theories.} Adv. Math. 137 (1998), no. 1, 1--81.

\bibitem{HoLySc1990}
Holm, D. D.; Lysenko, W. P.; Scovel, J. C. \textit{Moment invariants for
the Vlasov equation.} J. Math. Phys. 31 (1990), no. 7, 1610--1615.


\bibitem{HoRaTrYo2004}
Holm, D. D.; Ratnanather, T. J.; Trouv{\'e}, A.; Younes, L. {\it Soliton dynamics in computational anatomy.} Neuroimage 23 (2004) S170--S178.

\bibitem{HoSt03}
Holm, D. D.; Staley, M. F. \textit{Wave structure and nonlinear balances in a family of evolutionary PDEs.} SIAM J. Appl. Dyn. Syst. 2 (2003), no. 3, 323--380. 

\bibitem{HoTrYo2007}
Holm, D. D.; Trouv{\'e}, A.; Younes, L.
\textit{The Euler-Poincar\'e theory of metamorphosis}.
Preprint.

\bibitem{KoHaLi2001}
Koscielniak, S.; Hancock, S.; Lindroos, M. {\it Longitudinal holes in debunched particle beams in storage rings, perpetuated by space-charge forces.} Phys. Rev. ST Accel. Beams 4 (2001), no. 4, 044201.

\bibitem{KuMa}
Kupershmidt, B. A.; Manin, Ju. I. \textit{Long wave equations with a
free surface. II. The Hamiltonian structure and the higher equations.}
Funktsional. Anal. i Prilozhen. 12 (1978), no. 1, 25--37.

\bibitem{Ku2007}
P. A. Kuz'min, 
Two-component generalizations of the Camassa-Holm equation. 
{\em Math. Notes} {\bf 81} (2007), 130-134.
arXiv:math/0610786v2 [math.SG]

\bibitem{Le1979}
Lebedev, D. R. {\it Benney's long waves equations: Hamiltonian formalism.} Lett. Math. Phys. 3 (1979), no. 6, 481--488.

\bibitem{LeMa}
Lebedev, D.R.; Manin, Ju. I. \textit{Conservation laws and Lax
representation of Benney's long wave equations.} Phys. Lett. A 74 (1979),
no. 3-4, 154--156.

\bibitem{MaRa99}
Marsden, J. E.; Ratiu, T. S. {\it Introduction to mechanics and symmetry. A basic exposition of classical mechanical systems.} Second edition. Texts in Applied Mathematics, 17. Springer-Verlag, New York, 1999.

\bibitem{MaWe}
Marsden, J. E.; Weinstein, A. \textit{The Hamiltonian structure of the
Maxwell-Vlasov equations.} Phys. D 4 (1981/82), no. 3, 394--406.

\bibitem{MaWeRaScSp}
Marsden, J. E.; Weinstein, A.; Ratiu, T.; Schmid, R.; Spencer, R. G.
\textit{Hamiltonian systems with symmetry, coadjoint orbits and plasma
physics.} Proceedings of the IUTAM-ISIMM symposium on modern developments
in analytical mechanics, Vol. I (Torino, 1982). Atti Accad. Sci. Torino
Cl. Sci. Fis. Mat. Natur. 117 (1983), suppl. 1, 289--340.

\bibitem{MoBaJaLeNgShTa}
Moore, R.; Balbekov, V.; Jansson, A.; Lebedev, V.; Ng, K.Y.; Shiltsev, V.; Tan, C.Y. {\it Longitudinal bunch dynamics in the Tevatron.} Proceedings
of the Particle Accelerator
Conference 
Vol. 3 (2003), 1751--1753.

\bibitem{QiTa2004}
Qin, H.; Tang, W.M. {\it Pullback transformations in gyrokinetic theory.}
Phys. of Plasmas 11 (2004), no. 3, 1052--1063.

\bibitem{ScWe}
Scovel, C.; Weinstein, A. \textit{Finite-dimensional Lie-Poisson
approximations to Vlasov-Poisson equations.} Comm. Pure Appl. Math. 47
(1994), no. 5, 683--709.

\bibitem{WeMo}
Weinstein, A.; Morrison, P. J. \textit{Comments on:
``The Maxwell-Vlasov equations as a continuous Hamiltonian system''.}
Phys. Lett. A 80 (1980), no. 5-6, 383--386.

\bibitem{ScFe2000}
Schamel, H.; Fedele, R. {\it Kinetic theory of solitary waves on coasting beams in synchrotrons.} Phys. Plasmas 7 (2000), no. 8, 3421--3430. 

\bibitem{Venturini}
Venturini, M. {\it Stability analysis of longitudinal beam dynamics using noncanonical Hamiltonian methods and energy principles.} Phys. Rev. ST Accel. Beams 5 (2002), no. 5, 054403.

} 

\end{thebibliography}


\end{document}